\input harvmac
\noblackbox
\newcount\figno
\figno=0
\def\fig#1#2#3{
\par\begingroup\parindent=0pt\leftskip=1cm\rightskip=1cm\parindent=0pt
\baselineskip=11pt
\global\advance\figno by 1
\midinsert
\epsfxsize=#3
\centerline{\epsfbox{#2}}
\vskip 12pt
{\bf Fig. \the\figno:} #1\par
\endinsert\endgroup\par
}
\def\figlabel#1{\xdef#1{\the\figno}}
\def\encadremath#1{\vbox{\hrule\hbox{\vrule\kern8pt\vbox{\kern8pt
\hbox{$\displaystyle #1$}\kern8pt}
\kern8pt\vrule}\hrule}}

\input epsf

\overfullrule=0pt

\def\vps{\vert P \vert_s}
%
\def\cmt{\widetilde{\CM}}
\def\tilde{\widetilde}
\def\bar{\overline}

%
\def\inbar{\,\vrule height1.5ex width.4pt depth0pt}
\def\IB{\relax{\rm I\kern-.18em B}}
\def\IC{\relax\hbox{$\inbar\kern-.3em{\rm C}$}}
\def\ID{\relax{\rm I\kern-.18em D}}
\def\IE{\relax{\rm I\kern-.18em E}}
\def\IF{\relax{\rm I\kern-.18em F}}
\def\IG{\relax\hbox{$\inbar\kern-.3em{\rm G}$}}
\def\IH{\relax{\rm I\kern-.18em H}}
\def\II{\relax{\rm I\kern-.18em I}}
\def\IK{\relax{\rm I\kern-.18em K}}
\def\IL{\relax{\rm I\kern-.18em L}}
\def\IM{\relax{\rm I\kern-.18em M}}
\def\IN{\relax{\rm I\kern-.18em N}}
\def\IO{\relax\hbox{$\inbar\kern-.3em{\rm O}$}}
\def\IP{\relax{\rm I\kern-.18em P}}
\def\IQ{\relax\hbox{$\inbar\kern-.3em{\rm Q}$}}
\def\IR{\relax{\rm I\kern-.18em R}}
\font\cmss=cmss10 \font\cmsss=cmss10 at 7pt
\def\IZ{\relax\ifmmode\mathchoice
{\hbox{\cmss Z\kern-.4em Z}}{\hbox{\cmss Z\kern-.4em Z}}
{\lower.9pt\hbox{\cmsss Z\kern-.4em Z}}
{\lower1.2pt\hbox{\cmsss Z\kern-.4em Z}}\else{\cmss Z\kern-.4em
Z}\fi}
\def\IGa{\relax\hbox{${\rm I}\kern-.18em\Gamma$}}
\def\IPi{\relax\hbox{${\rm I}\kern-.18em\Pi$}}
\def\ITh{\relax\hbox{$\inbar\kern-.3em\Theta$}}
\def\IOm{\relax\hbox{$\inbar\kern-3.00pt\Omega$}}

\def\p{\partial}

\font\zfont = cmss10 

\def\bigone{\hbox{1\kern -.23em {\rm l}}}
\def\ZZ{\hbox{\zfont Z\kern-.4emZ}}

\def\half{{\litfont {1 \over 2}}}

\def\CM{{\cal M}}

\def\CS{{\cal S}}
\def\CN{{\cal N}}
\def\CX{{\cal X}}

\def\a{\alpha}
\def\b{\beta}
\def\d{\delta}

\def\IR{\relax{\rm I\kern-.18em R}}
\def\I1{\relax{\rm I\kern-.6em 1}}
\def\Dsl{\,\raise.15ex\hbox{/}\mkern-13.5mu D}
\def\Gsl{\,\raise.15ex\hbox{/}\mkern-13.5mu G}
\def\Csl{\,\raise.15ex\hbox{/}\mkern-13.5mu C}
\font\cmss=cmss10 \font\cmsss=cmss10 at 7pt
\def\pam{\partial_-}
\def\pap{\partial_+}
\def\pao{\partial_0}
\def\cp{{\cal P}}
\def\co{{\cal O}}
\def\cl{{\cal L}}
\def\CH{{\cal H}}

\font\zfont = cmss10 

\def\bigone{\hbox{1\kern -.23em {\rm l}}}
\def\ZZ{\hbox{\zfont Z\kern-.4emZ}}
\def\half{{\litfont {1 \over 2}}}

\def\CM{{\cal M}}

\def\IL{\relax{\rm I\kern-.18em L}}
\def\IH{\relax{\rm I\kern-.18em H}}
\def\IR{\relax{\rm I\kern-.18em R}}
\def\IC{\relax\hbox{$\inbar\kern-.3em{\rm C}$}}
\def\IZ{\relax\ifmmode\mathchoice
{\hbox{\cmss Z\kern-.4em Z}}{\hbox{\cmss Z\kern-.4em Z}}
{\lower.9pt\hbox{\cmsss Z\kern-.4em Z}}
{\lower1.2pt\hbox{\cmsss Z\kern-.4em Z}}\else{\cmss Z\kern-.4em
Z}\fi}
\def\CM {{\cal M}}
\def\CN {{\cal N}}
\def\CR {{\cal R}}
\def\CD {{\cal D}}
\def\CF {{\cal F}}

\def\CP {{\cal P }}
\def\CL {{\cal L}}
\def\CV {{\cal V}}

\def\CH {{\cal H}}
\def\CC {{\cal C}}

\def\CS {{\cal S}}

\def\CM {{\cal M}}
\def\CN {{\cal N}}

\def\CP {{\cal P }}

\def\CV{{\cal V }}

\def\CS {{\cal S }}

\font\manual=manfnt \def\dbend{\lower3.5pt\hbox{\manual\char127}}

\def\IZ{\relax\ifmmode\mathchoice
{\hbox{\cmss Z\kern-.4em Z}}{\hbox{\cmss Z\kern-.4em Z}}
{\lower.9pt\hbox{\cmsss Z\kern-.4em Z}}
{\lower1.2pt\hbox{\cmsss Z\kern-.4em Z}}\else{\cmss Z\kern-.4em
Z}\fi}
\def\half {{1\over 2}}

\def\p{\partial}

\def\CM {{\cal M}}
\def\CN {{\cal N}}

\def\CP {{\cal P }}

\def\CV{{\cal V }}

\def\CS {{\cal S }}


\def\IZ{\relax\ifmmode\mathchoice
{\hbox{\cmss Z\kern-.4em Z}}{\hbox{\cmss Z\kern-.4em Z}}
{\lower.9pt\hbox{\cmsss Z\kern-.4em Z}}
{\lower1.2pt\hbox{\cmsss Z\kern-.4em Z}}\else{\cmss Z\kern-.4em
Z}\fi}
\def\IB{\relax{\rm I\kern-.18em B}}
\def\IC{{\relax\hbox{$\inbar\kern-.3em{\rm C}$}}}
\def\ID{\relax{\rm I\kern-.18em D}}
\def\IE{\relax{\rm I\kern-.18em E}}
\def\IF{\relax{\rm I\kern-.18em F}}
\def\IG{\relax\hbox{$\inbar\kern-.3em{\rm G}$}}
\def\IGa{\relax\hbox{${\rm I}\kern-.18em\Gamma$}}
\def\IH{\relax{\rm I\kern-.18em H}}
\def\II{\relax{\rm I\kern-.18em I}}
\def\IK{\relax{\rm I\kern-.18em K}}
\def\IP{\relax{\rm I\kern-.18em P}}

\def\IQ{\relax\hbox{$\inbar\kern-.3em{\rm Q}$}}
\def\IP{\relax{\rm I\kern-.18em P}}

\def\inbar{\,\vrule height1.5ex width.4pt depth0pt}

\def\p{\partial}

\font\cmss=cmss10 \font\cmsss=cmss10 at 7pt
\def\IR{\relax{\rm I\kern-.18em R}}


\def\boxit#1{\vbox{\hrule\hbox{\vrule\kern8pt
\vbox{\hbox{\kern8pt}\hbox{\vbox{#1}}\hbox{\kern8pt}}
\kern8pt\vrule}\hrule}}
\def\mathboxit#1{\vbox{\hrule\hbox{\vrule\kern8pt\vbox{\kern8pt
\hbox{$\displaystyle #1$}\kern8pt}\kern8pt\vrule}\hrule}}


\def\inbar{\,\vrule height1.5ex width.4pt depth0pt}

\def\p{\partial}

\font\cmss=cmss10 \font\cmsss=cmss10 at 7pt
\def\IR{\relax{\rm I\kern-.18em R}}

\lref\maldacena{J. Maldacena,
``N=2 Extremal Black Holes and Intersecting Branes,''
 Phys.Lett. B403 (1997) 20-22; hep-th/9611163.}
\lref\cs{S. Cherkis and J. Schwarz, ``Wrapping the M-Theory
Five-Brane on K3,'' Phys. Lett. {\bf 403B} (1997) 225;
hep-th/9601029.}
\lref\sv{A. Strominger and C. Vafa, ``Microscopic Origin Of The
Bekenstein-Hawking Entropy,''Phys. Lett. {\bf 379B} (1996) 99;
hep-th/9703062.}
\lref\hpt{C. M. Hull, G. Papadopoulos, and P. K. Townsend,
``Potentials For $(p,0)$ and $(1,1)$ Supersymmetric Sigma Models
With Torsion,'' Phys. Lett. {\bf B316} (1993) 291.}
\lref\hp{P. Howe and G. Papadopoulos, ``Further remarks on the
geometry of two-dimensional non-linear $\sigma$-models, Class.
Quant. Grav. {\bf 5 } (1988) 1647. }
\lref\who{C. G.
Callan, Jr., J. A. Harvey, and A. Strominger, ``World-brane Actions
For String Solitons,'' Nucl. Phys. {\bf B367} (1991) 60, and
``Supersymmetric String Solitons,'' in the Proceedings, String
Theory and Quantum Gravity '91 (Trieste, 1991).}
\lref\adhm{M. F. Atiyah, V. G. Drinfeld, N. J. Hitchin, and Y. I. Manin,
``Construction Of Instantons,'' Phys. Lett. {\bf 65A} (1978) 185.}
\lref\millermoore{S.D. Miller and G. Moore,
``Landau-Siegel zeroes and black hole entropy,''
hep-th/9903267}
\lref\arthatt{G. Moore, ``Arithmetic and attractors,'' hep-th/9807087;
``Attractors and arithmetic,'' hep-th/9807056.}
\lref\corrigan{E. Corrigan and P. Goddard,
``Construction Of Instanton And Monopole Solutions And Reciprocity,''
Annals of Physics {\bf 154} (1984) 253.}
\lref\strom{A. Strominger, ``Heterotic Solitons,'' Nuclear Physics
{\bf B343} (1990) 167.}
\lref\harv{C. G. Callan, Jr., J. A. Harvey, and A. Strominger,
``Worldsheet Approach To Heterotic Instantons and Solitons,''
Nucl. Phys. {\bf B359} (1991) 611.}
\lref\kron{P. Kronheimer, ``The Construction Of ALE Spaces As
Hyper-K\"{a}hler Quotients,'' J. Diff. Geom. {\bf 29} (1989) 665.}
\lref\hw{C. Hull, E. Witten, ``Supersymmetric sigma models and the
heterotic string,''
Phys. Lett. {\bf 160B}  (1985), 398.}
\lref\ckp{P. Claus, R. Kallosh, A. Van Proeyen, ``M5-brane and
superconformal (0,2) tensor multiplet in 6 dimensions,''
hep-th/9711161.}
\lref\pt{G. Papadopoulos, P. Townsend,
``Massive sigma-models with (p,q) supersymmetry,''
Class. Quant. Grav. {\bf 11 } (1994) 515.}
\lref\witten{E. Witten, ``Sigma models and the ADHM construction of
instantons,'' hep-th/9508039, J. Geom. Phys. {\bf 15}  (1995) 215.}
\lref\lambert{N. D. Lambert, ``Quantizing the (0,4) Supersymmetric
ADHM Sigma Model,'' Nucl. Phys. {\bf B460} (1996) 221,
hep-th/9508039.}
\lref\bbs{K. Becker, M. Becker and A. Strominger, ``Fivebranes,
Membranes and Non-Perturbative String Theory,''  Nucl. Phys. {\bf
B456} (1995) 130, hep-th/9507158.}
\lref\msw{J. Maldacena, A. Strominger and E. Witten, ``Black Hole Entropy
in M Theory,'' J. High Energy Phys. 12 (1997) 002, hep-th/9711053.}
\lref\fiveac{I. Bandos et al, ``Covariant Action for the
Super-Five-Brane of M-Theory,''
Phys. Rev. Lett. {\bf 78} (1997) 4332; hep-th/9701149.}
\lref\cfg{S. Cecotti, S. Ferrara  and L. Girardello, ``Geometry of
Type-II Superstrings and the Moduli of Superconformal Field
Theories,''  Int. J. Mod. Phys. {\bf A4}(1989) 2475.}
\lref\mdi{R. Dijkgraaf and G. Moore, ``Balanced Topological Field
Theories'',
hep-th/9608169. }
\lref\mns{M. Douglas and G. Moore, ``D-branes, Quivers, and ALE
Instantons,'' hep-th/9603167\semi G. Moore, N. Nekrasov and S.
Shatashvili, ``Integrating Over Higgs Branches,'' hep-th/9712241.}
\lref\hklr{N.~Hitchin, A.~Karlhede, U.~Lindstrom, and M.~Rocek,
``Hyper-K\"{a}hler metrics and supersymmetry,'' Commun. Math. Phys.
{\bf 108} (1987) 535.}
\lref\wittfive{E. Witten, ``Five-Brane Effective Action In M-Theory,''
hep-th/9610234.}
\lref\ccdf{A.C.~Cadavid, A.~Ceresole, R.~D'Auria and S.~Ferrara,
``11-Dimensional Supergravity Compactified on Calabi-Yau
Threefolds'', Phys. Lett. {\bf B357} (1995) 76, hepth/9506144\semi
I.~Antoniadis, S.~Ferrara and T.R.~Taylor, ``N=2 Heterotic
Superstring and its Dual Theory in Five Dimensions,''Nucl. Phys.
{\bf B460} (1996) 489, hep-th/9511108.}
\lref\fkm{S.~Ferrara, R.R.~Khuri
and R.~Minasian ``M-Theory on a Calabi-Yau Manifold'' Phys. Lett.
{\bf B375} (1996) 81-88; hep-th/9602102.}
\lref\sch{M. Aganagic, J. Park, C. Popescu, J. Schwarz, ``World-Volume
Action of the M Theory Five-Brane,'' hep-th/9701166\semi J.
Schwarz, ``Coupling a Self-Dual Tensor to Gravity in Six
Dimensions,'' hep-th/9701008.}
\lref\pst{P. Pasti, D. Sorokin, M. Tonin, ``On Lorentz-invariant
actions for chiral p-forms'', hep-th/9611100.}
\lref\kod{K. Kodaira, ``Complex  manifolds and deformation of
complex structures,'' Springer-Verlag,  New York, 1985.}
\lref\agmv{L. Alvarez-Gaume, G. Moore, C. Vafa, ``Theta functions,
modular invariance and strings'', Commun. Math. Phys. {\bf 106}
(1986) 1. }
\lref\hast{ J. A. Harvey, and A. Strominger,
``The Heterotic String is a Soliton,'' Nucl. Phys. {\bf B449}
(1995) 535, hep-th/9504047.}
\lref\rd{R. Dijkgraaf, ``Instanton Strings and Hyper-K\"ahler Geometry,''
hep-th/9810210.}
\lref\nr{K. Narain, M. Sarmadi and E. Witten, ``A note on toroidal
compactification of heterotic string theory'', Nuclear Physics {\bf
B279} (1987) 369\semi P. Ginsparg ``On toroidal compactifications
of heterotic superstrings'', Phys. Rev. D. {\bf 35} (1987) 648\semi
G. Cardoso, D. Lust and T. Mohaupt ``Threshold corrections and
symmetry enhancement in string compactifications'',
hep-th/9412209.}
\lref\hmm{J.A. Harvey, R. Minasian and G. Moore, ``Non-abelian
Tensor-multiplet Anomalies,'' J. High Energy Phys. 09(1998) 004;
hep-th/9808060.}
\lref\mm{R. Minasian and G. Moore, ``K-theory and Ramond-Ramond
charge,''   JHEP 11 (1997) 002; hep-th/9710230.}
\lref\wittenk{E. Witten, ``D-branes and K-theory,''
hep-th/9810188.}
\lref\douglas{M.R. Douglas and B. Fiol,
``D-branes and discrete torsion II,'' hep-th/9903031.}
\lref\yz{S.T. Yau and E. Zaslow, ``BPS States, String Duality,
and Nodal Curves on K3,'' Nucl.Phys. {\bf B471} (1996) 503;
hep-th/9512121.}
\lref\hm{J.A. Harvey and G. Moore, to appear.}
\lref\vhs{P.A. Griffiths, `` Periods of integrals on algebraic
manifolds, III,'' Publications I.H.E.S. {\bf 38} (1970) 125\semi W.
Schmid,
``Variation of Hodge structure: The singularities of the period mapping,''
Inv.
Math. {\bf 22} (1973) 211 \semi P. Griffiths and W. Schmid, ``Recent
developments in Hodge Theory: A discussion of techniques and results,'' in
{\it
Discrete subgroups of Lie groups and applications to moduli} Bombay
Colloquium
1973, OUP 1975 \semi J. Carlson, M. Green, P. Griffiths, and J. Harris,
``Infinitesimal variations of Hodge structures,'' Compositio Mathematica
{\bf
50} (1983) 109 \semi see also ``Topics in transcendental algebraic
geometry,''
ed. P.A. Griffiths, Princeton University Press, 1984 \semi J.-L. Brylinski
and
S. Zucker, ``An overview of recent advances in Hodge theory,'' in {\it
Several
Complex Variables VI}, W. Barth and R. Narasimhan eds. Springer Verlag
1990
\semi V.S. Kulikov and P.F.
Kurchatov, ``Complex algebraic varieties: Periods of integrals and Hodge
Structures,'' in {\it Algebraic Geometry III}, eds. A.N. Parshin and I.R.
Shafarevich, Springer-Verlag,  New York, 1998}
\lref\fr{D. S. Freed, ``Special K\"{a}hler Manifolds'', hep-th/9712042.}
\lref\kallosh{K. Behrndt, G. Lopez Cardoso, B. de Wit, R. Kallosh,
D. Lust and T. Mohaupt, ``Classical and quantum N=2 supersymmetric
black holes,'' Nucl. Phys. {\bf B488} (1997) 236; hep-th/9610105.}
\lref\kawaimohri{T. Kawai and K. Mohri,
``Geometry of $(0,2)$ Landau-Ginzburg orbifolds,''
 Nucl. Phys. {\bf B425} (1994) 191; hep-th/9402148.}
\lref\mms{J. Maldacena, G. Moore, and A. Strominger,
``Counting BPS black holes in toroidal type II string theory,''
hep-th/9903163.}
\lref\mmtt{R. Minasian, G. Moore, A. Todorov and D. Tsimpis,
in progress}
\lref\open{A. Strominger, ``Open $p$-branes," Phys. Lett. {\bf 383B}
(1996) 44; hep-th/9512059.}
\lref\calabi{E. Calabi, ``Kahler metrics and holomorphic vector
bundles," Ann. Sci. Ecole Norm. Sup. {\bf 12} (1979) 269.}
\lref\morrison{D. Morrison, Trieste Conference 1985}

%

\Title{\vbox{\baselineskip12pt
\hbox{YCTP-P11-99}
\hbox{IASSNS-99/44}
\hbox{hep-th/9904217}
}}
{\vbox{
\centerline{Calabi-Yau black holes}
\smallskip
\centerline{and}
\smallskip
\centerline{ (0,4) sigma models }
}}

\bigskip
\centerline{Ruben Minasian$^1$, Gregory Moore$^{1,2}$ and Dimitrios
Tsimpis$^1$}
\bigskip
\centerline{$^1$Department of Physics, Yale University}
\centerline{New Haven, CT 06520, USA}
\medskip
\centerline{$^2$ School of Natural Sciences, Institute for Avanced Study}
\centerline{Olden Lane, Princeton, NJ 08540, USA}

\bigskip
\centerline{\bf Abstract}
\medskip

\noindent
When an $M$-theory fivebrane wraps a holomorphic surface $\CP$ in a
Calabi-Yau 3-fold $X$ the low energy dynamics is that of a black
string in 5 dimensional $\CN=1$ supergravity.
 The infrared dynamics on the
string worldsheet is an  $\CN = (0,4)$ 2D
conformal field theory.
Assuming the 2D CFT can be described as a nonlinear
sigma model, we describe the target space geometry
of this model in terms of the data of $X$ and $\CP$.
Variations of weight two Hodge structures enter
the construction of the model in an interesting way.

\Date{April 30, 1999}


\newsec{Introduction}

D-brane and M-brane
models of black holes have provided an
extremely intriguing approach to an understanding
of black hole entropy \sv\
and promise to lead to important insights
in other aspects of black hole physics.

The program of Strominger and Vafa
is based on mapping the low
energy dynamics of certain configurations
of branes to the  conformal field theory
of an effective string. The  derivation of this
conformal field theory is best understood
(and already quite subtle)
for black holes in backgrounds preserving
16 supersymmetries, such as IIB compactification
on $K3 \times S^1$. In this paper we will
investigate an analogous conformal
field theory for 4D black holes in
backgrounds with 8 unbroken supersymmetries.

Specifically, in
this paper we continue the investigation
of the microscopic dynamics of wrapped
five-branes following the work of
Maldacena, Strominger,
and Witten  \msw. We consider
$M$-theory compactifications
on $\IR^{1,3}\times S^1 \times X$ where $X$ is a
 nonsingular compact Calabi-Yau 3-fold.
The radius of $S^1$ is taken to be large with respect to the
length scale set by  $X$,
which is in turn large compared to the 11D Planck scale.
We usually will take the background $3$-form $C^{(3)}$ to vanish. If  an
$M5$-brane worldvolume $W_6$ wraps $\IR\times S^1\times \CP$,
where $\CP$ is
 a  four-manifold $\CP \subset X$
then the
resulting object is a string $\CS$ with
worldsheet $W_2$ wrapping $\IR\times S^1$
 in $\IR^{1,3}\times S^1$.
At long distances the supergravity background
is that of a black
  hole in $\IR^{1,3}$  with $8$ unbroken
supersymmetries at infinity. The 4-manifold
$\CP$ must be a
 holomorphically
embedded complex surface  to
preserve supersymmetry. In this case
the low energy dynamics of the string  $\CS$
is described by a $(0,4)$ CFT
and the  number of massless
boson and fermion degrees of freedom
can be expressed purely
in terms of the topology of $\CP$ and
of  its embedding into $X$ \msw.

Knowing the number of
massless degrees of freedom
suffices to determine the
entropy microscopically, but for many
purposes one would certainly
 like to know the data
of the
$(0,4)$ conformal field theory of
$\CS$ in much more detail. The object
of this paper is to express this  data
 in terms of the data of
the ambient Calabi-Yau geometry and
the topology and geometry of $\CP$.

In this introduction we summarize
the structure of the sigma model
that we will find. The  detailed
justification is described in the
subsequent sections. Much of what
we say is implicitly (and sometimes
explicitly) described in \msw.
Let us begin with the overall count of
the degrees of freedom.
In a supersymmetric configuration
the surface $\CP$ is a
divisor for a holomorphic line bundle $\CL$ over $X$. Let $P=
[\CP]\in H^2(X;\IZ)$ be the first Chern class of $\CL$. It is
Poincar\'e dual to the 4-cycle defined by $\CP$. Macroscopically,
in the 5D supergravity obtained from $M$ theory on $X$, the string
is a black string and $P$ is  the charge of the string.
Using index theory and holomorphic geometry \msw\ computed
the  left- and right-moving
central charges:
\eqn\cntr{\eqalign{c_R &
= 6D + \half c_2 \cdot P\cr
c_L &=  6D + c_2 \cdot P, \cr }}
where $D:={1 \over 6} \int_{X}P^3$ and $c_2 \cdot P := \int_X P \,
c_2(TX)$,
verifying microscopically the entropy
computed macroscopically in
\kallosh.
 These central charges can also be obtained
from the requirement of the complete anomaly cancellation in
five dimensions \hmm.

We now describe some of the local geometry
of the target space. This is obtained by
considering the collective coordinates
of the wrapped 5brane. These include
the collective coordinates associated to the
5 scalars $X^a$ and
the chiral two-form $\beta$ of the
5-brane worldvolume tensormultiplet.

We begin with the collective coordinates $\varphi$
associated to the five scalars.
The space of supersymmetric
wrappings of charge $P$ is the set of divisors
in $X$ in the class $P$. This is called a
``linear system'' because all divisors are
zero-loci of global holomorphic
sections of $\CL$, and the latter is a linear
space. Of course two sections related by
a multiplicative constant have the same
divisor, so the linear system
  is  just a  projective space
\eqn\linsys{
\vert P \vert := \IP H^0(\CP, \CL\vert_{\CP})
= \IC\IP^N.
}
Assuming
$\CP$ is a smooth ample divisor, as we should to apply
classical geometry \msw,
the Riemann-Roch formula gives the dimension
of the linear system \linsys:
\eqn\dimels{
N := D+ {1\over 12} c_2\cdot P - 1.
}
Taking into account the position in noncompact
$\IR^3$, the target space of the scalars is
\eqn\scalars{
\varphi: W_2 \rightarrow \IR^3 \times
\vert P\vert.
}

In this paper we will make the important
restriction that the
fivebrane wraps a {\it smooth} 4-cycle $\CP$.
Thus if
$\CD$ is the discriminant locus of
singular divisors in
the linear system we  restrict to maps
$\varphi$ with image in
\eqn\psmooth{
\vert P\vert_s:= \vert P \vert -\CD.
}
Moreover, because of monodromy, we will
even restrict
attention to maps into a
 local neighborhood $\CU\subset \vps$.
\foot{We thank E. Witten for stressing the
importance of the monodromy.}

Now we consider the collective coordinates
arising from the chiral two-form $\beta $
on worldvolumes of the form $W_6=W_2 \times \CP$.
The massless modes are associated with
harmonic two-forms on $\CP$. Since the form
is chiral there are
$b_2^- := b_2^-(\CP)$ left-moving and $b_2^+:=b_2^+(\CP)$
right-moving chiral bosons. Moreover, as shown in \msw,  one can
express these topological invariants of $\CP$ in terms of $D$ and
$c_2\cdot P$:
\eqn\bpm{
\eqalign{
b_2^- & =4 D + {5\over 6} c_2\cdot P - 1\cr
b_2^+ & = 2 D + {1\over 6} c_2\cdot P - 1\cr}
}

If $b_1(\CP)=0$ (which follows if $b_1(X)=0$) the
only fermions are rightmoving. These pair
up to form
\eqn\rightmult{
 D + {1\over 12} c_2\cdot P
}
$\CN = (0, 4)$ scalar multiplets with
both left- and right- moving scalars.
In addition there are
real purely leftmoving scalars
neutral under supersymmetry.
Since
\eqn\nbee{
N = D + {1\over 12} c_2\cdot P -1  = \half(b_2^+-1)
}
there are
$\vert \sigma(\CP)\vert = b_2^-(\CP) - b_2^+(\CP)$
such scalars.

Let us now consider the scalars from the chiral
two-form in more detail. The splitting into
 left-movers and right-movers follows from the decomposition
\eqn\sdasd{
H^2(\CP;\IR) =  H^{2,-}(\CP,\IR)\oplus  H^{2,+}(\CP;\IR)
}
into anti-self-dual and self-dual parts respectively.
Since $\CP$ is K\"ahler we may further decompose
\eqn\toozer{
H^{2,+}(\CP;\IR) = [H^{2,0}(\CP) \oplus H^{0,2}(\CP)]_{\IR}
\oplus \IR\cdot J
}
where $J$ is the Kahler class of $\CP$ induced by
that of $X$,
while
$H^{2,-}(\CP;\IR)$ is purely of Hodge type $(1,1)$.
A crucial point is
that the splitting \sdasd\toozer\
depends on $\CP$ (i.e., on the values of $\varphi$),
and hence
on the (weight two) Hodge structure of $H^2(\CP;\IZ)$.
Thus, a natural framework for working with the
$(0,4)$ model is the theory of variation of
Hodge structures.
(See the references in \vhs\ for some
useful background material.)

The Hodge structure on $H^2(\CP)$ decomposes into a
``fixed part'' and a ``variable part'' (as
functions of $\CP$):
\eqn\fxvr{
H^2(\CP;\IR) = H^2_f(\CP;\IR) \oplus  H^2_v(\CP;\IR)
}
where $H^2_v$ is the orthogonal complement of
$H^2_f$ in the Hodge metric $(\theta_1, \theta_2):=
\int_\CP \theta_1 \wedge \theta_2$.
The ``fixed'' or ``rigid''  part is simply the space of 2-forms
which extend to $X$. As pointed out in
\msw, since $\CP$ is ample the restriction map
\eqn\restr{
\iota^*: H^2(X,\IZ)
\rightarrow H^2(\CP;\IZ)
}
is injective so $H^2_f(\CP;\IR) \cong
H^2(X;\IR)$.
Physically, the splitting \fxvr\ means the
$(0,4)$ sigma model splits
(up to possible discrete identifications by
a finite group) into a product of two
sigma models which we call the ``universal
factor'' and the ``entropic factor.'' The
terminology refers to the intuition that
$P$ should be regarded as large, so that
$D$ is a very large positive integer, determining
the leading term in the
black hole entropy.

The CFT for the universal
factor is easily described. It consists of
a single  $(0,4)$ multiplet with target space
$\IR^3 \times S^1$ (for left- and right-movers)
together with
$h^{1,1}(X)-2$ purely leftmoving bosons.
\foot{Except when  $h^{1,1}(X)=1$. In this
case  the compact scalar in
$S^1$ is purely right-moving and there
is no left-moving gauge bundle.}
 Since $\beta$ can be shifted by
large gauge transformations in $H^2(X;\IZ)$
the universal factor is just a $(0,4)$
Narain model with leftmoving gauge group
of rank $h^{1,1}(X)-2$. The Narain data
is obtained from the projection of
$H^2(X;\IZ)\otimes \IR $
onto the definite signature subspaces of
the
quadratic form $(\theta_1, \theta_2)=
\int_{\CP}\theta_1 \wedge
\theta_2 =  \int_X P \wedge \theta_1 \wedge
\theta_2$.

The entropic factor  is more subtle and is
the focus of much of this paper. In
describing this model we will
make the important assumption
that the model can be described by a
geometrical Lagrangian (see footnote 1  above).
Roughly speaking,  a $(0,4)$
 sigma model Lagrangian is determined
by a choice of  target
space $\cmt$  which has
 a hyper-K\"ahler
 connection (with torsion)  together with
 a triholomorphic vector
bundle with connection
$\CV \rightarrow \cmt$.
The detailed conditions on the
sigma model Lagrangian are written
in equations 2.13 - 2.20 below.

In the following sections
we  will argue  that in our case
the target space $\cmt$ of the
entropic factor has  a
holomorphic projection
\eqn\intsys{
p: \widetilde{\CM} \rightarrow \IP^N
}
where the
 projective space $\IP^N$ is the linear
system $\vert P \vert$. Physically, the degrees of freedom
describing the fibers of $p$ have their origin in  those of the
self-dual two-form $\beta$ on $W_6$. The fibers are complex tori of
complex dimension $N= \half (b_2^+-1)$. Moreover, the vector bundle
$\CV$ has
real  rank $\vert \sigma(\CP)\vert - (h^{1,1}(X)-2)
= b_2^-(\CP) - b_2^+(\CP) - (h^{1,1}(X)-2)$,
and the connection on $\CV$ is vertical,
and flat.

In section 5.3 below we argue that the
torsion of the connection on $T\cmt$
is zero, so that the  metric on $\cmt$ is hyper-K\"ahler.
This metric  may be described as
follows. The  metric of  the ambient Calabi-Yau $X$ induces a
metric on the normal bundle of $\CP\subset X$, and therefore a
metric on the linear system $\vert P \vert$. This metric is
K\"ahler,
\foot{Warning: the metric on the
linear system is {\it not} the
Fubini-Study metric.}
 and, by
a version of the Calabi ansatz/c-map (\calabi/\cfg) there is an
induced hyper-K\"ahler metric on $T^* \vert \CP \vert$. Thus, the
the local geometry of $\widetilde{\CM}$ is that of $T^* \vert P
\vert$ with a hyper-K\"ahler structure.

Having described the local geometry of the target space we
now turn to global issues. There are several interesting issues one
should address, but we focus on only one, namely the nature of the
fibers of $p$ (working locally in a patch of $\vert P \vert$).
The derivation of the sigma model Lagrangian uses
 the chiral fivebrane Lagrangians of  \refs{\pst, \sch}.
Unfortunately, the   formalism of these papers does not determine
the way in which zeromodes of the $b_2^-$
 left-chiral and $b_2^+$  right-chiral
scalars are paired with each other.
Therefore we must resort to  some guess-work.

The key motivation for our guess is that
we expect that the target space
of the sigma model $\cmt$ should be
{\it compact}. Otherwise it is hard to
understand how we could have
 finite dimensional spaces of
BPS states. Since the linear system
$\vert P \vert$ is compact
the  question
reduces to compactness of the fibers.
Therefore,
all the $b_2(\CP)$ modes due to the
chiral $\beta$-field should be compact
scalars. The  most
natural way to achieve this is to
assume that the fiber above $\CP$ is
just a conformal field theory on a  torus
with a flat connection, i.e., a Narain model.
The data of the Narain model consists of a
choice of lattice $\Gamma$ of signature
$(p,q)$ of zeromodes of scalars and
an orthogonal projection of $\Gamma\otimes \IR$
to the definite signature subspaces
defining the spectrum of left- and right-moving
parts of the winding/momentum lattice.
In our case we will take
$\Gamma=H^2(\CP;\IZ)$
and  the
orthogonal projection is
\eqn\orthogproj{
H^2(\CP; \IZ) \otimes \IR \rightarrow
H^{2,+}(\CP; \IR)\perp H^{2,-}(\CP; \IR)
}
induced from the metric on $\CP$.
Thus, according to our hypothesis, the
fibers of the projection in \intsys\ are
complex tori of dimension $N$, so we have
a holomorphic integrable system.
Two interesting subtleties in this
discussion are,
first, there is a nontrivial flat connection
on the toroidal fibers, and second,
by a mechanism mentioned in \msw\ most of
the charges in $H^2(\CP;\IZ)$ are not
conserved (we comment on this briefly in
section 6).

Finally we mention another motivation for the present work. D-brane
models of black holes appear to have interesting arithmetic
properties \refs{\arthatt, \millermoore}, at least for the case of
black holes in backgrounds with 16 supersymmetries. One can
entertain various conjectures about the arithmetic nature of
D-brane black holes in Calabi-Yau compactification and several of
these are related to questions about the  numbers of BPS states
$\dim \CH_{BPS}(\gamma)$ for charge $\gamma \in H^{\rm even}(X)$.
The results of the present paper might  help to elucidate the
nature of these BPS degeneracies. Our hope is that these
degeneracies can be studied using $(0,4)$ elliptic genera.

We summarize the remaining sections as follows.
In section two we review the general form of
$(0,4)$ Lagrangians. In section three we
review the
relation between unbroken supersymmetry
and holomorphically wrapped 4-cycles.
In section four we describe in detail the
derivation of the collective coordinates
and the relation to variation of Hodge
structures. We derive carefully the
$(0,4)$ supermultiplets by reducing the
supersymmetry transformations of the
6D tensormultiplet. In section five
we use the Kaluza-Klein ansatz of
section four in the chiral 5-brane
action and derive the geometry of the
target space of the $(0,4)$ model.
In section six we comment on some
of the global aspects of the target
space model. These involve the
toroidal fibers of \intsys\ and
their relation to Narian models.
In section seven we describe
briefly what we think are some
of the most interesting open
problems raised by this paper.
Many conventions and technical
points may be found in the appendices.
Appendix D describes the close
analogy of the models in this
paper with the strings obtained
by wrapping D3 branes around
holomorphic curves in a K3 surface.

\newsec{Geometrical data for $(0,p)$  $\sigma$-models}

In this section we summarize the geometrical
data used to construct $(0,4)$ supersymmetric
Lagrangians. This material is standard, and
this  section follows mostly \refs{\hw, \pt}.

In two spacetime
dimensions the supersymmetry algebra of type $(0,p)$ is carried by
$p$ negative-chirality supersymmetries $Q^I_-$, $I=1,...p$,
obeying:
\eqn\salg{\{Q^I_-,Q^J_-\}=2\delta^{IJ}P_-}
In the construction of the $(0,4)$ sigma-models it is convenient to
consider a formulation with only $(0,1)$ manifest supersymmetry.
$(0,1)$ superspace consists of two Bose coordinates $x^+,x^-$ and a
single negative-chirality anticommuting coordinate $\theta^-$. Our
sigma-model is defined by a  map from (0,1)
superspace $\Sigma$ to a $d$-dimensional target manifold ${\cal
M}$, given by scalar superfields ${\Phi^i(x,\theta^-), i=1,...d}$.
In general there can also be another field which is a
 section of the vector bundle $S_-\otimes\phi^{*}{\cal V}$ over
$\Sigma$, given by negative-chirality \foot{the subscript refers to
the fact that in a free theory the $\theta$-independent component
in the expansion of $\Lambda$ would be a left-moving fermion in the
sense that $\partial_-\lambda^a_+=0$} spinor superfields
${\Lambda^a_+(x,\theta^-), a=1,...n}$, where $n$ is the fibre
dimension of ${\cal V}$ and $S_-$ is the spinor bundle over
$\Sigma$. ${\cal V}$ is equipped with a positive definite metric
$h_{ab}$ and a connection $A_i{}^a{}_b$ with curvature
$F_{ij}{}^a{}_b$ valued in some subgroup of $O(n)$. The requirement
of (0,4) SUSY imposes additional constraints which will be analyzed
at the end of this section. The superfields have the following
expansions:
\eqn\comps{\eqalign{\Phi^i(x,\theta^-)&=\phi(x)^i+i\theta^-
\psi^i_-(x) \cr
\Lambda^a_+(x,\theta^-)&=\lambda^a_+(x)+i\theta^-F^a(x)
\cr}}
The action for the model in terms of $(0,1)$ superfields reads:
\eqn\superaction{\eqalign{ S=\int d^2x d\theta^- &\Biggl(
(g_{ij}(\Phi)+b_{ij}(\Phi))D_-\Phi^i\partial_+\Phi^j  \cr
 &+i\Lambda^a_+(D_- \Lambda^b_+ +D_- \Phi^i A_i{}^b{}_c
\Lambda^c_+)h_{ab}+imC^a\Lambda^bh_{ab} \Biggr) \cr}}
where
\eqn\deq{D_-={\partial \over \partial \theta^-}+i \theta^-
{\partial \over \partial x^-}}
where $A_i{}^a{}_b$ is the connection on  ${\cal V}$ with a
curvature $F_{ij}{}^a{}_b$. After eliminating the auxiliary fields
and expanding in components, \superaction\ reads:
\eqn\compaction{\eqalign{S=\int & d^2x \Biggl[(g_{ij}+b_{ij})
\partial_+\phi^i \partial_- \phi^j+ig_{ij} \psi^i_-\nabla^{(+)}_+
\psi^j_- -i\lambda^a_+{\cal D}_-\lambda^b_+h_{ab} \cr
-&{1 \over 2} \lambda^a_+ \lambda^b_+ \psi^i_- \psi^j_-
F_{ijab}+m \nabla_i C^a \psi^i_
- \lambda^b_+h_{ab}-
{1 \over 4} m^2 C^a C^b h_{ab} \Biggr] \cr}}
where $\nabla^{(\pm)}$ is the covariant derivative with respect to
the connection with torsion
\eqn\connection{\Upsilon^{(\pm)i}{}_{jk} = \Gamma^i{}_{jk}\pm H^i{}_{jk}}
\eqn\torsion{H_{ijk}= {3 \over 2} \partial_{[i}b_{jk]}}
and
\eqn\covarder{{\cal D}_-\lambda_+=\partial_-\lambda^a_+
+\partial_-\phi^iA_i{}^a{}_b\lambda^b_+}

The action \superaction\ is manifestly invariant under
supersymmetry transformations
\eqn\susytr{\eqalign{\delta \Phi^i &=i\eta_+ D_- \Phi^i  \cr
\delta \Lambda^a_+ &=-i\eta_+ D_- \Lambda^a_+ .\cr}}
Let us now assume that \superaction\ possesses additional
supersymmetries, parametrized by anticommuting parameters
${\eta^r_+,\,r=1,...p-1}$, of the form
\eqn\adsusytr{\eqalign{\delta \Phi^i &=i\eta^r_+
J_r{}^i{}_j (\Phi) D_- \Phi^j  \cr
\delta \Lambda^a_+ &=-\delta \Phi^i  A_i{}^a{}_b\Lambda^b_+
+\eta^r_+ I_r{}^a{}_b(\Phi) S^b -m \eta^r_+ t^a_r(\Phi) \cr}}
where
\eqn\es{S^a=2\nabla_- \Lambda^a_+ +mC^a.}
and $J_r, I_r, t_r$ are to be determined.

Invariance of the action under \susytr\ and \adsusytr, and on-shell
closure of the supersymmetry algebra are equivalent to the
following set of conditions \refs{\pt, \hpt, \hp}
\eqn\cnda{J_rJ_s=-\delta_{rs}+f_{rs}{}^tJ_t}
\eqn\cndb{N(J_r,J_s)^i_{jk}=0}
\eqn\cndc{F_{ij}J_r{}^i{}_{[k}J_s{}^j{}_{l]}=F_{kl}\delta_{rs}}
\eqn\cndd{J_r{}^k{}_{(i}g_{j)k}=0}
\eqn\cnde{\nabla^{(+)}J=0}
\eqn\cndf{I_r{}^c{}_{(a}h_{b)c}=0}
\eqn\cndg{\partial_{i}(t_r^aC^bh_{ab})=0}
\eqn\cndh{{\nabla}_it^a_r=J_r{}^j{}_i{\nabla}_j C^a}
Here $N$ is essentially the Nijenhuis tensor. The SUSY algebra
closes on-shell on $\lambda^a_+$ by virtue of
\cnda, \cndb, \cndc, \cndh.

Conditions \cnda\ - \cndf\ can be summarized as follows \pt: ${\cal
M}$ admits three complex structures obeying the algebra of
quaternions, the metric on ${\cal M}$ is hermitian with respect to
all three complex structures and the holonomy of the connection
$\Upsilon^{(+)}$ is a subgroup of $Sp(d/4)$. The bundle ${\cal
V}\otimes {\bf C}$ is holomorphic with respect to all three complex
structures and carries an Hermitian metric $h_{ab}$.

\newsec{Supersymmetrically wrapped fivebranes}

The low energy states of the black string $\CS$ are obtained from
small deformations of supersymmetrically wrapped cycles. Thanks to
the existence of a $\kappa$-symmetric fivebrane action we can
analyze which configurations preserve supersymmetry, following the
analysis of \bbs.

The  unbroken supersymmetries are the result of combining the
supersymmetry $\delta_{\epsilon}\Theta
=\epsilon$ with $\kappa$-symmetry $\delta_{\kappa}\Theta =
2(1+{\bar \Gamma})\kappa$, where the matrix ${\bar \Gamma}$ is
field dependent (the explicit expression can be found in \fiveac )
and has the property that $(1 \pm {\bar \Gamma})$ are projection
operators. The  condition of unbroken supersymmetry is
\eqn\unbroken{(1-{\bar \Gamma})\epsilon \equiv P_- \epsilon = 0.}

Let the $M5$-brane be stretched in the $X^{0} - X^5$ directions,
and consider a compactification on a Calabi-Yau threefold along
$X^2,..., X^7$. We consider the wrapping of the fivebrane on a
four-cycle $\cal P$ stretched in $X^2-X^5$  directions. The
coordinates on the fivebrane world-sheet are denoted by
$\sigma^{\alpha},\,\,\;
\alpha=0,...,5$. It is convenient to choose a gauge such that
$\sigma^0=X^0,\,\,\,\sigma^1=X^1$. Also let $X^m,X^{\bar{m}},\,\,\,
m=1,2$ be a complex basis for $X^2-X^5$ (choosing a gauge such that
$dX^6+idX^7$ is an eigenvector of the complex structure on the
Calabi-Yau). An eleven-dimensional spinor $\epsilon$ is decomposed
as
\eqn\elevspin{\epsilon_{\pm} = \lambda^{(3)} \otimes
\lambda^{(2)}_{\pm} \otimes \xi^{(6)}}
where $\lambda^{(2)}_{\pm}$ is a spinor
of $Spin(2)_{01}$ of positive (negative)
chirality, $\lambda^{(3)}$ is a
$Spin(3)_{8910}$ spinor and $\xi^{(6)}$
is the covariantly constant spinor of
the Calabi-Yau. The eleven-dimensional
$\Gamma$-matrices can be decomposed as
follows
\eqn\elevgamma{\eqalign{&\Gamma^{8,9,10} = \gamma^{8,9,10}
\otimes\rho^{(2)}_{01} \otimes\rho^{(6)}  \cr
&\Gamma^{0,1}={\I1}_3 \otimes \gamma^{0,1} \otimes\rho^{(6)}  \cr
&\Gamma^{2,3,4,...,7}={\I1}_5 \otimes \gamma^{2,3,4,...,7}  \cr}}
where $\rho^{(2)}_{01}=i\gamma^0\gamma^1$ is the chirality operator
of $Spin(2)_{01}$ and $\rho^{(6)}$ is the chirality operator acting
on the Calabi-Yau spinors. We let
$\Gamma_{M_1...M_k}=\Gamma_{[M_1}...\Gamma_{M_k]}$.
Denoting  a positive (negative)- chirality spinor on the Calabi-Yau
by $\xi^{(6)}_{\pm}$, we can choose a normalization such  that the
following identities hold
 \eqn\cychrlspin{\gamma_{\bar{m}}\xi^{(6)}_+=0,\,\,\,\,\,\,\,
\gamma_{\bar{n}
pq}\xi^{(6)}_+=2iJ_{\bar{n}[p}\gamma_{q]}\xi^{(6)}_+ }
where $J$ is the K\"ahler form for $X$.
Also,
we  have passed to a complex basis for the $\gamma$ matrices such
that $\gamma^1_c=
\gamma^2 +i\gamma^4 = (\gamma^{\bar 1}_c)^{\dagger}$ and  $\gamma^2_c=
\gamma^3
+i\gamma^5 = (\gamma^{\bar 2}_c)^{\dagger}$. We will omit the
subscript on the complex matrices whenever there is no possibility
of confusion. Equation
\cychrlspin\ implies
\eqn\gmttf{\gamma_{\bar{m}\bar{n}pq}\xi^{(6)}=
(J_{\bar{n}p}J_{\bar{m}q}-J_{\bar{m}p}J_{\bar{n}q})\xi^{(6)}}
where we used $J_{\bar{m}n}=ig_{\bar{m}n}$ and the anticommutation
relations.
We thus have
\eqn\pim{\eqalign{P_-\epsilon_{\pm}
&={1\over 2}\left( 1\pm \partial_2 X^{\bar{m}} \partial_3 X^{\bar{n}}
\partial_4 X^{p}
\partial_5 X^{q} {1 \over 4} \left( J_{\bar{n}p}J_{\bar{m}q}-(\bar{m}
\leftrightarrow  \bar{n}) \right) \right)  \epsilon_{\pm} \cr
}}
where $\partial_i=\partial/\partial\sigma^i$, $\, A_i=2,...,5,\,\,$
and we have used $\gamma^{\bar{1}\bar{2}12}=-4\gamma^{2345}$. As
expected,  $P_-\epsilon_{\pm}=0$ implies
\eqn\zf{(dV)_4=\pm \iota^{*}({1 \over 2}J\wedge J)}
where $dV_4$ is the volume form of the part of the fivebrane
wrapping the Calabi-Yau four-cycle and $\iota^{*}(J\wedge J)$ is
the pullback of $J\wedge J$ to the fivebrane.

{}From \pim, \zf\ it follows that for holomorphic (anti-holomorphic)
cycles
$\cal
P$ only $\epsilon_+$ ($\epsilon_-$) can satisfy $P_-\epsilon=0$. Thus only
one
eighth of the supersymmetry is preserved and the resulting $\sigma$-model
is
chiral $(4,0)$ ($(0,4)$)

\newsec{The massless $(0,4)$ supermultiplets }

We now turn to the description of the massless degrees of freedom
on $\CS$ arising from small fluctuations around a supersymmetric
wrapping of the fivebrane worldvolume on $\IR\times S^1  \times
\CP$. Our method will be to  reduce the six-dimensional $(2,0)$ multiplet
along $\CP$. Of course, this presupposes some facts about the
equations of motion. These will be justified in the section five.

\subsec{Reduction of the scalars}

The five scalars of the 6D
 $(2,0)$ tensormultiplet, denoted
$X^a$ (cf $B.2$) parametrize the position of the fivebrane in
eleven dimensions. When we wrap an M-theory fivebrane on
a real  four-cycle $\CP$ inside a Calabi-Yau manifold, three scalars
(call them
$X^{8,9,10}$) parametrize the position of the string in the
noncompact
dimensions, and the remaining two
describe  the
position of the cycle $\cal P$ inside the Calabi-Yau and
should therefore  be thought of infinitesimally
as sections of the normal
bundle.

The massless scalars arise from deformations of
the position of $W_2 \times \CP$ preserving unbroken
supersymmetry. The space of deformations of $\cal P$ as a
complex submanifold of $X$ has tangent space
\eqn\tansp{
T_\CP \vert P \vert = H^0(\CP,\CN)
}
where $\CN$ is the normal bundle. We  may
identify $\CN \cong \CL\vert_\CP$.
It follows from the index theorem that
if $\CP$ is ample
the  complex  dimension of the space of deformations
of $\CP$ is $N= D + {1
\over 12}c_2P - 1$.

When we consider small fluctuations of the wrapped fivebrane on a
cycle $\CP$ we are really considering a family of cycles
$\cp_{\varphi}$ near  a point $\CP_{\varphi= 0}$ in the moduli
space of deformations of $\CP$. These fit into a holomorphic
fibration
\eqn\family{
\matrix{ \CX&  \cr \downarrow& \pi  \cr \CU & \cr}
}
where $\CU\subset \vps$ is a neighborhood of $\varphi=0$. The fibers
of this family are diffeomorphic to $\CP_{\varphi=0}$,
 but have variable complex
structure.
In  appendix $C$ we show
that there is an injection
\eqn\eqvdfs{
0 \rightarrow H^0(X,{\cal L})/{\IC}\cdot s \cong  H^0 ({\cal P},
{\cal L}|_{\cp})
\rightarrow H^{0,1}( T{\cal P} ) }
where $\CP$ is the vanishing locus of the section $s$.
By Kodaira-Spencer theory $H^{0,1}(T\CP)$ is
  the space of infinitesimal
deformations of the complex structure of ${\cal P}$ \kod.
Thus, a first order deformation of $\varphi$ induces
a nonzero deformation of complex structure. This
will be important in the next section.

Finally, we write out the Kaluza-Klein ansatz for the
scalars describing fluctuations around $\CP$.
Choosing a basis,
$\upsilon_I,$ for  $H^0\left( \cp, {\cal L}|_{{\cal P}} \right)$
and a coordinate system on $X$ so that
$dX^6 + i dX^7$ is normal to $\CP$
we may expand, to first order in $\varphi^I$,
\eqn\phxpn{X^6+iX^7=\upsilon_I \varphi^I }
to obtain complex two-dimensional massless fields $\varphi^I$,
$I =1,...,N$.These
scalars are both left- and right-moving.

\subsec{Reduction of the worldvolume two-form}

The chiral two-form $\beta$ on $W_6$ reduces to
left-moving and right-moving scalars according
to the decomposition into self-dual and anti-self
dual parts as in equations \sdasd\toozer\ of the
introduction.
As mentioned in the introduction, $H^2(\CP)$ carries
a polarized Hodge structure of weight two.
The Hodge decomposition is simply
\eqn\hgstr{
H^2(\CP;\IZ)\otimes \IC = H^{2,0}(\CP) \oplus H^{1,1}(\CP)
\oplus H^{0,2}(\CP)
}
and the polarization is given by the Hodge metric.

\goodbreak\midinsert
\centerline{\epsfxsize 3.5truein\epsfbox{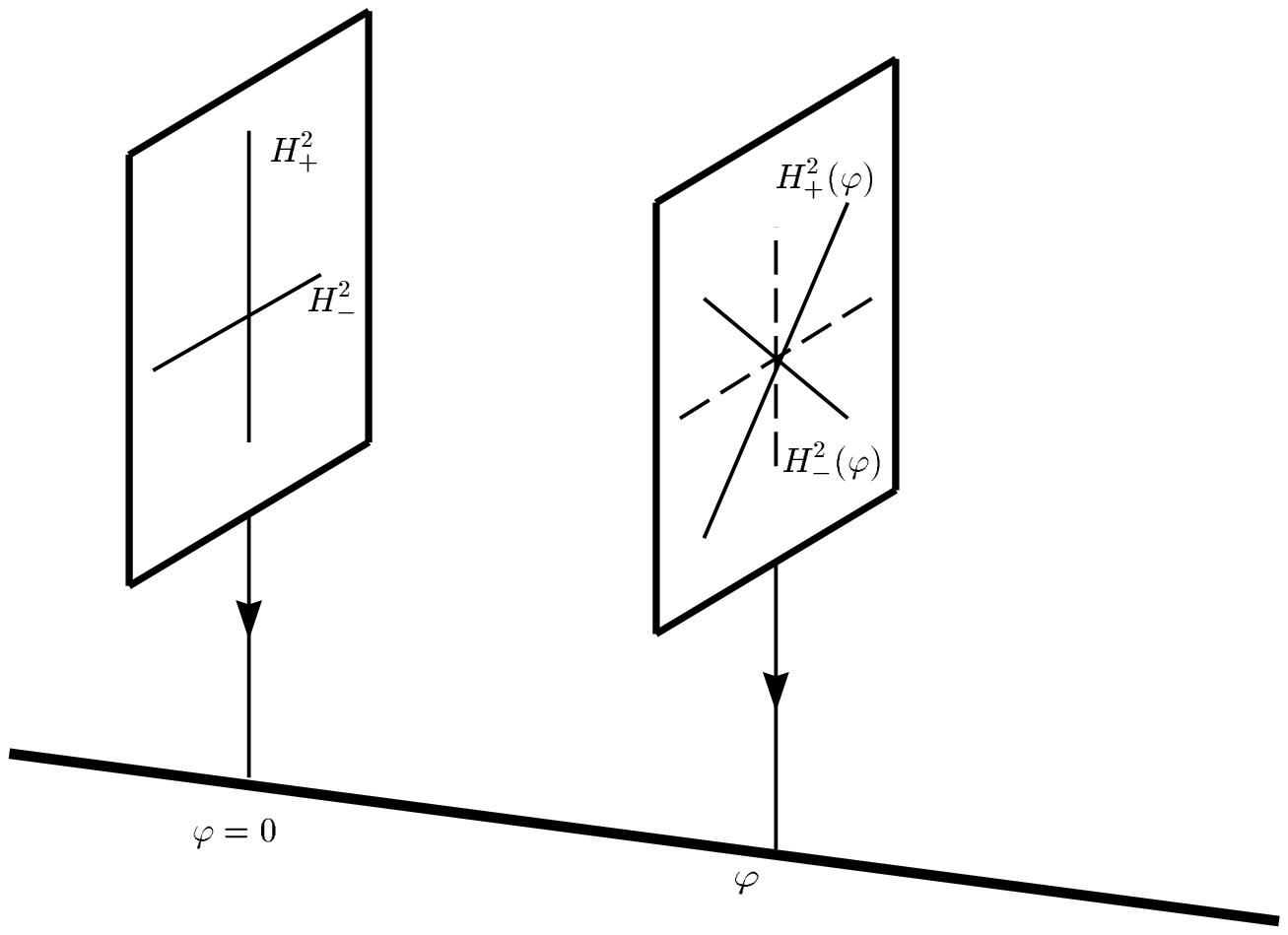}}
\leftskip 2pc \rightskip 2pc
\noindent{
\ninepoint\sl \baselineskip=8pt
{\bf Fig.1} As we move in the moduli
space of deformations, we may
use $\CC^\infty$ diffeomorphisms to
define a flat
 basis for $H^2(\cp;\IZ)$ in the
fibers. However both the Hodge
decomposition and the decomposition into
self-dual/antiself-dual parts change.}
\smallskip\endinsert

It is important to bear in mind that the decomposition
\hgstr\ and hence the decomposition of
$\beta$ into left- and right-moving scalars
depends on $\varphi$. As $\varphi$ changes the
K\"ahler metric and complex structure on $\CP_\varphi$
varies, as illustrated in Fig. 1. This is the
standard geometrical realization of variation of
Hodge structures \vhs. Since the fibers are all diffeomorphic
we can choose a family of $\CC^\infty$ diffeomorphisms
and define a local system (i.e. a  flat bundle) with
a locally constant basis for $H^2(\CP;\IZ)$. Extending by
linearity defines the
 Gauss-Manin connection on the bundle $R^2\pi_*(\IC)$
over $\CU$ whose fiber at $\CP$ is $H^2(\CP)$.
The Gauss-Manin connection
 allows us to differentiate in
the $\varphi^I$ direction, and is crucial in
deriving the low-energy $(0,4)$ Lagrangian.

If we choose a
smoothly-varying basis
 $\omega_I$, $I=1,\dots, \half(b_2^+-1)$
 of harmonic $(2,0)$ forms on $\CP$, and
 $\omega_{-a}, \,\, a=1,..., b_2^-({\cal P})$,  of anti-selfdual
$(1,1)$-forms on $\CP$,
then these bases will rotate into one another
in accord with Griffiths tansversality. That is, if we
define the holomorphically
varying filtration $H^2(\CP) = \CF^0 \supset \CF^1\supset \CF^2$
by
\eqn\filtration{
\eqalign{
\CF^0 & = H^{2,0} \oplus H^{1,1}\oplus H^{0,2} \cr
\CF^1 & = H^{2,0} \oplus H^{1,1} \cr
\CF^2 & = H^{2,0} \cr}
}
then
\eqn\omshift{\nabla: \CF^p \rightarrow \CF^{p-1} \otimes\Omega^1(\CU).}
Or, in plain english, the connection matrix is upper triangular and
increases $p$ in the decomposition into $(p,q)$ forms by at most
one. We can split the Hodge structure into a fixed and a
variable part as in \fxvr\ of the introduction.
 The Hodge structure $H^2_f$ is fixed (as a function of $\CP$)
because it is purely of type $(1,1)$ for
all $\CP$.

Finally, we write out the Kaluza-Klein ansatz for the
chiral two-form $\beta$ as:
\eqn\bdcmp{
\beta = \rho^a \omega_{-, a  } + 4 (\pi^I \omega_I + c.c.)
+  u_4 J
}
The two-dimensional complex scalars $\pi^I$ and the real scalar
$u_4$ are right-moving whereas the $b_2^-$
 real scalars $\rho^a$ are left-moving.

\subsec{Reduction of $(2,0)$ tensormultiplet fermion fields along  $\cal
P$}

We now describe the Kaluza-Klein ansatz for the fermions in the
6D tensormultiplet on $W_2 \times \CP$. We will expand these
in terms of harmonic $2$-forms and $0$-forms on $\CP$. The
conceptual reason we can do this is the following.

The 5-brane breaks local 6D Lorentz symmetry on $W_6$
as
\eqn\subgroup{
Spin(1,1)\times Spin(4) \hookrightarrow  Spin(1,5)
}
So the fermions are in the ${\bf 4 = (+\half; 2,1)
\oplus (-\half; 1,2)} $. Moreover, the tensormultiplet
theory has a $Spin(5) = USp(4) $ $\CR$ symmetry group
from local Lorentz rotations in the normal directions
(in 11D)
to $W_2\times \CP$. The Calabi-Yau background breaks this
to $Spin(3) \times Spin(2)$, where $Spin(3)$ are rotations
in the noncompact normal directions and $Spin(2)$ is the
structure group of the normal bundle $\CN$ for $\CP$ in $X$.
The restriction of the spin bundle on $X$ to $\CP$ decomposes
as:
\eqn\decspin{
S^+(TX)  \cong S^+(T\CP) \otimes K^{1/2}\oplus S^-(T\CP) \otimes K^{-1/2}
}
but, because $\CP$ is K\"ahler,
\eqn\spinfrm{
S^+(T\CP) \otimes K^{1/2} \cong \Omega^{0,0}(\CP) \oplus
\Omega^{2,0}(\CP).}
Hence we expand the zeromodes in terms of $0$-forms and $2$-forms
on $\CP$.

In order to reduce the supersymmetry transformations we will
need to make \spinfrm\ more explicit.
Our conventions for
 six-dimensional supersymmetry  are  in  Appendix B.
We choose a
basis for six-dimensional $\Gamma$ matrices to be
\eqn\sxmgamma{\eqalign{\Gamma^{0,1} &= \gamma^{0,1}
\otimes\rho^{(4)}  \cr
\Gamma^{2,3,4,5} &= {\I1}_2 \otimes \gamma^{2,3,4,5}  \cr}}
where
\eqn\gmzroonetfv{\eqalign{\gamma^0=i\sigma^2&;\,\,\,\,\,
\gamma^1=\sigma^1 \cr
\gamma^{2,3,4} = \pmatrix{
0 & \sigma^{1,2,3} \cr
\sigma^{1,2,3} & 0 \cr
}&; \,\,\,\,\,  \gamma^{5} = \pmatrix{
0 & i \, \I1_2 \cr
-i \, \I1_2 & 0 \cr
}
\cr}}

We will decompose the covariantly constant spinor on the Calabi-Yau   into
two- and four-dimensional parts
\eqn\cyspin{\xi^{(6)}= \xi^{(2)} \otimes \xi}
We take $\xi^{(6)}$ to be anti-chiral in order to conform to the
chirality of the tensor multiplet.  Note that the $Spin(2)_{67}$
and $Spin(4)_{2345}$ spinors $\xi^{(2)}$ and $\xi$ are not
covariantly constant but only projectively covaraintly constant.
That is, they are parallel up to a phase, and the phase
cancels between $\xi^{(2)}$ and $\xi$.

Again, using \subgroup,
the 6D tensormultiplet spinors
$\psi^{(6)}$ in their turn decompose as
\eqn\tzspin{\psi^{(6)}_i = \psi^I_{i-} \otimes \Delta^I_{(i)} +
\psi^0_{i-} \otimes \xi_{(i)}. }
Here $i=1,2,3,4$ is a $USp(4)$ $\CR$-symmetry index
(see appendix B) and there is no summation on $i$. Moreover,
\eqn\defdelta{ \Delta^I_{(i)} = \cases{
\omega^I_{\bar{mn}} \gamma^{\bar{mn}} \xi  & for $i=1,3$ \cr
\omega_{I,mn} \gamma^{mn} \xi^* & for $i=2,4$
}}
and
\eqn\defxi{ \xi_{(i)} =\cases{
\xi  & for $i=1,3$ \cr
\xi^* & for $i=2,4$}}
This decomposition is consistent with the symplectic reality condition,
which with our conventions reads
\eqn\sympre{\psi_{1-} = i \psi_{2-}^{\dagger},
\,\,\,\,\,\,\, \psi_{3-} = -i \psi_{4-}^{\dagger}}

Thus the two-dimensional spinors $\psi^I_{i-}$ correspond to two
complex spinor
degrees of freedom; they carry a subscript minus  since
only right-moving spinors survive the wrapping as massless
degrees of freedom.
Note, for use in section 5.2 that since
$\gamma_{\bar{m}} \xi^{(6)} =0$
one learns that
$\gamma_{\bar{m}} \xi =0$.

\subsec{Reduction of the supersymmetry transformations}

We now come to the reduction of the 6D supersymmetry transformations. In
principle, we should take into account all the complications of kappa
supersymmetry and the exact superisometries unbroken by the background
determined by $X$ and $\CP$. However, as explained in section 4.6 below,
it is
sufficient for our purposes to consider the reduction of the supersymmetry
transformations from flat space.

As a preliminary to the calculation
 it will prove very useful to
relate the basis $\upsilon_I$ of \phxpn\
(used for the scalars) to the basis $\omega_I$
of  holomorphic
$(2,0)$ forms on $\CP$.
Abstractly this is a
consequence of
\eqn\convvv{H^0(\cp, \CL\vert_{\CP} ) \cong H^0(\cp,
\Omega^2_{\cp}) \cong H^{2,0}(\cp),} which follows from the
adjunction formula $(K_X \otimes \CL )|_{\cp} \cong K_{\cp}$ and
the triviality of the canonical bundle $K_X$ of $X$.
More explicitly we have a
 relation between the basis of the holomorphic
two-forms on $\cal P$, $\omega_{I,mn},$ and
$\upsilon_I$ given by:
\eqn\convert{\upsilon_I =
\xi^{\dagger} \omega_{I,mn} \gamma^{mn} \xi^{*}; \,\,\,\,\,
\bar{\upsilon}^I = \xi^{Tr} {\bar \omega}^{I,mn} \gamma_{mn} \xi}
Conversely, one can write
$\iota(\upsilon_I)\Omega|_{\cp}^{(3,0)}
=
\omega^{(2,0)} _I$ where $\iota$ is a contraction and
$\Omega^{(3,0)}$ is a nowhere zero
 holomorphic three-form on $X$

Substituting the above expansions into the supersymmetry
transformations of the tensormultiplet
the  supersymmetry
transformation of the $\beta$-field yields
\eqn\redtensor{\eqalign{\delta \pi^I &=
-2( \epsilon_{i+} )^{\dagger} (1+ iT)_{ij}  \psi^I_{j-}\cr
\delta u^4 &= 2 \left( \epsilon_{i+}  \right)^{\dagger}
T_{ij}  \psi^0_{j-} \cr
\delta \rho^a &= 0 \cr} }
where $T = \hat{\gamma}^6 \hat{\gamma}^7$ and $\half(1+iT)$ is a
projection operator. Similarly, the reduction of the susy
transformations for the $X^a$ gives
\eqn\redscals{\eqalign{\delta \varphi^I &=
2(\epsilon_{i+})^{\dagger} \left(
\hat{\gamma}^6(1+iT ) \right)_{ij}  \psi^I_{j-}\cr
\delta X^{8,9,10} &= 2 \left( \epsilon_{i+}  \right)^{\dagger}
\hat{\gamma}_{ij}^{8,9,10}\psi^0_{j-} \cr}}
The second lines in
\redtensor\ and \redscals\ can be joined into a single equation
\eqn\transun{\delta u^A = 2  \left( \epsilon_{i+}
\right)^{\dagger}  \check{\gamma}^A_{ij} \psi^0_{j-} }
where $A=1,...4$ and we have defined $\check{\gamma}^{A=1,2,3} =
\hat{\gamma}^{8,9,10} $
and $ \check{\gamma}^{A=4} = T$, and
\eqn\defyou{
u^1=X^8,\qquad   u^2=X^9, \qquad u^3=X^{10} }

Finally, the transformation laws of the fermions reduces to
\eqn\fermred{\eqalign{\delta \psi^I_{2-} =
{1 \over 2} \partial_{-} \left( \varphi^I
\bar{\epsilon}_{4+}  + \pi^I \epsilon_{2+} \right);   \,\,\,\,\,\,
\delta \psi^I_{4-} &= {1 \over 2}
\partial_{-} \left(- \varphi^I \bar{\epsilon}_{2+} +
\pi^I \epsilon_{4+} \right)  \cr
\delta \psi^0_{i-}  ={1 \over 2}  \partial_{-} u_A \check{\gamma}^A_{ij}
\epsilon_{j+} \cr }}
where $\bar{\epsilon} = -i \epsilon^{\dagger},
\,\,\,\, \partial_{-}=- \partial_0 + \partial_1$. In
the first line of  \fermred\ we have found it more convenient to
work explicitly with components. Of course the other two
transformations ($\delta \psi^I_{1-}, \,\, \delta \psi^I_{3-}$) are
related to the above by the symplectic reality condition \sympre.
We will give a considerably more attractive form of this equation
in the next section.

\subsec{Assembling the multiplets and the  Hyper-K\"{a}hler structure}
We will now summarize the previous sections by
describing  the $(0,4)$ supermultiplets in terms of the
a hyper-K\"{a}hler structure.

Note first that from the above supersymmetry
transformations  the four real scalars $u_A$
and the four real component  spinor $\psi^0$ transform
amongst themselves. Since this multiplet is
present no matter what the topology of $X$ or
$\CP$ is,  we refer to it as  the  ``universal''
multiplet. It has some nice analogies with the
universal hypermultiplet of Calabi-Yau compactification
of type II strings.

We now cast the susy transformations in
 quaternionic form. By letting $w=u_4-iu_3; \,\,\,\, z=-u_2+iu_1$
the bosonic coordinate {\bf X} has the form
\eqn\bosqut{
{\bf X} = \pmatrix{ z & \bar w \cr - w & \bar z\cr}
}
Moreover if  we define the quaternions
\eqn\dfqtrthps{ {\bf  \Theta} = \pmatrix{
\eta & \bar{\theta} \cr
-\theta & \bar{\eta} \cr
} ; \,\,\,\,\, {\bf \Psi}^0 = \pmatrix{
\chi & \bar{\psi} \cr
-\psi & \bar{\chi} \cr
}   }
where we have set $\eta =i \epsilon_{2+}; \,\,\,\, \theta =
 -i \epsilon_{4+}; \,\,\,\, \chi =
\psi^0_{1-}; \,\,\,\, \psi = \psi^0_{4-}$,
then \transun\ reads
\eqn\qtruniv{\delta {\bf X} = -4i {\bf  \Theta} \cdot {\bf  \Psi}. }
The corresponding fermionic transformation (the second line of \fermred)
becomes
\eqn\qtrfrmrd{ \delta {\bf \Psi}^{\dagger} =-{i \over 2} \partial_-
{\bf X } \cdot {\bf \Theta}  }
These are the supersymmetry transformations of the universal
superfield in a manifestly $(0,4)$ invariant form.
 Geometrically, the four scalars parametrize
${\cal U} = {\IR}^3 \times S^1$
(the $u_4$ scalar coming from tensor field is periodic).

Similarly, if we define
\eqn\lowerpi{
\delta {\bar \pi}_I : = D_{I {\bar J}} \delta {\bar \pi}^{\bar J}
}
and
\eqn\defdij{
D_{I \bar J} := \int_{\CP} \omega_I \wedge
\bar \omega_{\bar J}
}
we see from the supersymmetry transformations that the  scalars
$\pi^I$ and $\varphi^I$ mix under $\delta^2$. Proceeding in the
same manner we define
\eqn\dfqtrdio{ \delta {\bf  X}^I=\pmatrix{
\delta \varphi^I & \delta \bar{\pi}_I  \cr
- \delta \pi^I & \delta \bar{\varphi}_I \cr
}; \,\,\,\,\,  {\bf  \Psi}^I=\pmatrix{
\chi^I & \bar{\psi}_I  \cr
-\psi^I & \bar{\chi}_I \cr
}   }
and
\eqn\thtqtrdf{ {\bf  \Xi} = \pmatrix{
\zeta & \bar{\lambda} \cr
-\lambda & \bar{\zeta} \cr
} }
where now $ \zeta=\epsilon_{4+}; \,\,\,\, \lambda =
\bar{ \epsilon}_{2+}; \,\,\,\, \chi^I =
\psi^I_{2-}; \,\,\,\, \psi^I = \psi^I_{4-}$.
We see that the transformations $\d \varphi^I$ and $\d {\bar
\pi}_I$ are joined together since the first lines of \redtensor\
and
\redscals\ can be expressed compactly as
\eqn\compactexpr{ \delta {\bf X}^I =
 -4i {\bf  \Xi} \cdot {\bf  \Psi}^I }
Similarly the fermionic transformations
(the first line of \fermred) become
\eqn\becomes{ \delta {\bf \Psi}^{I} =
{1 \over 2} {\bf \Xi}^{\dagger} \cdot \partial_- {\bf X }^I
}
And once more the $(0,4)$ supersymmetry is manifest. It is clear
from our construction that ${\bar \pi}_I$ play the role of
coordinates on the cotangent space. Thus, the target space is
locally $T^*\vert P
\vert$.

\subsec{Comparison with standard $(0,4)$ models}

Now that we have determined the field content and
$(0,4)$ multiplets let us begin to
 make contact with the general form of the
Lagrangian described in section 2. Since the target
space is locally $\CU \times T^*\vert \CP\vert$
there must be $N+1 = D+{1\over 12} c_2\cdot P$
 scalar multiplets. The scalars in these
multiplets have {\it both} left- and right-moving degrees of
freedom, whereas the complex scalars $\pi^I$ derived from \bdcmp\
are purely right-moving. On the other hand, the chiral 2-form
$\beta$ also gives $b_2^-$ real left-moving scalars $\rho^a$. Since
the signature of $\CP$ is negative, we must pair $b_2^+$ degrees of
freedom from the $\rho^a$ with the rightmovers $\pi^I$. The
remaining $\vert \sigma(\CP)\vert$ degrees of freedom correspond to
the left-moving fermions denoted by  $\lambda$ in section 2. We
will discuss this pairing of left- and right-moving degrees of
freedom further in section 5 below.

We can also compare with
  the supersymmetry transformations
of section two.
Let $Z^M=(X,\theta)$ be the superembedding
coordinates and let $M=(m,\mu); \,\,
A=(a,
\a)$ be ``curved'' and  ``inertial'' eleven dimensional indices respectively.
Lower-case
Latin (Greek) letters denote bosonic (fermionic) components. Up to local
Lorentz
transformations the vielbein transforms under supercoordinate
transformations as
$\delta_{\epsilon}E^A_M={\cal D}_M \epsilon^A$ where ${\cal D}_M$ is the
covariant derivative. We see that the fermionic transformations which
preserve a
given background are parametrized by covariantly constant spinors
$\epsilon$.
The fermionic symmetries of the fivebrane are:
\eqn\psym{\delta_{\epsilon}Z^M=E^M_{\a} \epsilon^{\alpha}; \,\,\,
\delta_{\kappa}Z^M
=(1+{\bar \Gamma})_{\beta}{}^{\a}\kappa^{\beta}E^M_{\alpha}}
where ${\bar \Gamma}$ is as in section 3.

In order to recover the field content of the six dimensional
$(2,0)$ tensor multiplet we need to do some gauge fixing. Upon
reducing the resulting equations to two dimensions we get highly
non-linear expressions. Ultimately we want to compare with the
supersymmetries of the sigma model presented in section 2 (which is
quadratic both in derivatives and in the right-moving fermions).
Such a truncation brings us back to the supersymmetry
transformations of appendix B.2 and their reduction, equations
(\compactexpr, \becomes). Comparing with
\adsusytr\ we can
 read off the three complex structures encoded in these equations: Let
us
choose a real basis $\delta \varphi^I = \delta \phi^{1I}+ i\delta
\phi^{2I};
\,\,\, \delta \pi^I:= D^{I \bar{J}} \delta \pi_{\bar{J}} =\delta
\phi^{3I}+ i \delta
\phi^{4I}$. Similarly
$\chi^I =\theta^{1I}+ i \theta^{2I}; \,\,\,  \psi^I =-\theta^{4I}+
i\theta^{3I}$ and $\zeta= \epsilon^0 +i\epsilon^1; \,\,\, \lambda=
\epsilon^2 +i\epsilon^3$. With these definitions \compactexpr\
takes the form
\eqn\rdftf{\delta \phi^{iI} = \epsilon^0 \theta^{iI}+ \epsilon^r
J_r{}^{iI}{}_{jJ}
\theta^{jJ}, \,\,\, i=1,\dots 4}
where
\eqn\cplxstrs{J_1=[\sigma^3 \otimes i\sigma^2]^i{}_j \delta^I{}_J, \,\,
J_2=[i\sigma^2
\otimes
\I1_2]^i{}_j \delta^I{}_J,
\,\, J_3=[\sigma^1 \otimes i\sigma^2]^i{}_j \delta^I{}_J}
One can verify that they satisfy $J_rJ_s=-(\delta_{rs} +
\varepsilon_{rst}J_t)$.

\newsec{Local target-space geometry }
The low-energy two-dimensional Lagrangian encodes  the geometry of
the target space of the $(0,4)$ model.
We will derive this Lagrangian by Kaluza-Klein
reduction of the chiral fivebrane Lagrangian of \refs{\pst, \sch}.

\subsec{Bosonic Lagrangian}
Let
us start by reducing the bosonic part of the five-brane action
presented in \sch.
The   action possesses manifest general
coordinate covariance only along five of the six worldvolume
dimensions. Here we will take the distinguished direction to be the
spatial direction of the two-dimensional world-sheet  $W_2$
 which is taken to be flat.
For our conventions/definitions we refer to appendix $A.2$.

The action consists of three terms
\eqn\lmbda{ L_1 = - \sqrt{ -det \left( G_{ \hat{\mu} \hat{ \nu} } +
G_{ \hat{\mu}  \rho } G_{\hat{\nu}  \lambda } \tilde{ H }^{\rho
\lambda } / \sqrt{ -G_5 } \right) } }
\eqn\lmbdb{ L_2 = -{1 \over 4} \tilde{ H }^{\mu \nu }
\p_1 \beta_{ \mu \nu } }
\eqn\lmbdc{ L_3 = {1 \over 8 } \varepsilon_{ \mu \nu \rho \kappa
\lambda } {G^{ 1 \rho } \over G^{11} } \tilde{ H }^{\mu \nu }
\tilde{ H }^{ \kappa \lambda  } }
We will use the ``static gauge'' $X^{\hat{\mu}} =
\sigma^{\hat{\mu}}$ in which the above expressions simplify and we
recover the field content of the $(2,0)$ six-dimensional multiplet
discussed in
the appendix B (note however that here we are using a gauge in which the
antisymmetric tensor is effectively five-dimensional in the sense that
$\b_{1
\hat{\mu} } = 0$). Moreover we will keep only terms at most quadratic in
$\p X$
and/or $H$. As explained before, when reducing to $W_2 \times {\cal P} $
the
only nonvanishing components of the field $\b$ are along ${\cal P}$. Hence
\lmbda\  - \lmbdc\ read
\eqn\lmbdaa{ L_1 =  {1 \over 2} g_{ab}\p_{\hat{\mu}}
X^a \p^{\hat{\mu}} X^b - {1 \over 4 }
\tilde{ H}^{\mu \nu } \tilde{ H }_{\mu
\nu } (1 + g_{ab}\p_{\rho} X^a \p^{\rho} X^b )
- {1 \over 2 }
\tilde{ H }^{\mu \kappa } \tilde{ H }_{\kappa}{}^{ \nu }g_{ab}
 \p_{\mu}
X^a \p_{\nu} X^b  }
\eqn\lmbdab{ L_2 = -{1 \over 8}
\varepsilon^{ABCD} \p_0 \b_{AB} \p_1 \b_{CD}  }
\eqn\lmbdac{ L_3 = {1 \over 4 } \tilde{ H }^{\mu \nu }
 H_{  \rho \mu \nu } g_{ab}\p^{ \rho } X^a \p_1 X^b }
where $g_{ab}$ is the metric on the space transverse to the
fivebrane. Keeping only up to two-derivative terms and dropping
terms with derivatives along $\cp$ (which are suppressed by the
size of $\cp$) the above expressions simplify further
\eqn\simplesch{ \eqalign{ S = \int_{W_6} dV \Biggl( {1 \over 2} &
\left( -\p_{0} X^a \p_0 X^b + \p_1 X^a \p_1 X^b \right) g_{ab}+ {1 \over
4} g^{AC}g^{BD}\p_0 \b_{AB} \p_0 \b_{CD} \cr
 &-{1 \over 8} \varepsilon^{ABCD} \p_0 \b_{AB} \p_1 \b_{CD} \Biggr), \cr}
}
Reducing the kinetic term of the scalars
in
\simplesch\ using
\phxpn\ we see that it gives rise to a term,
\eqn\curlf{\int_{W_2} d^2 \sigma
 \p_{(+} \varphi^I
\p_{-)} {\bar \varphi^{\bar J}}  G_{I {\bar J}} ,}
where $G_{I {\bar J}} = \int_{\cal P}dV G({\cal L})
\upsilon_I {\bar \upsilon_{\bar J}}, $ and $G({\cal
L})$ is the hermitian metric  on the line bundle ${\cal L}|_{\cal
P}
= N({\cal P}
\hookrightarrow X)$ (in complex notation, it has only one
component). By using a Fierz identity (see $B.1$) we can establish
the metric on the space of $\varphi$'s in terms of the intersection
matrix
\eqn\mercf{ G_{I {\bar J}}  =
D_{I {\bar J}} =
\int_{\cal P}  \, \omega_I \wedge  \bar{\omega}_{\bar J}.}

As for the chiral two-form,
the  reduction of \simplesch\ gives (omitting the universal
superfield)
\eqn\fixedact{S_0 = \int_{W_2}d^2 \sigma ( \p_{(0} {\bar \pi}_I  \p_{+)}
\pi_{\bar J}
 D^{I {\bar J}}  -
\pao \rho^a \pam \rho^b D_{ab}).}

\subsec{Fermionic Lagrangian}

Equations \curlf, \mercf\ and \fixedact\ contain some of the
essential information we need to extract the geometric data for the
$(0,4)$ Lagrangian. However, to extract all the data we must
consider the quadratic terms in fermions.
Therefore, we  look at the terms
quadratic in the right-moving fermions
containing exactly one derivative along
$W_2$. These  come from
the reduction to $W_2$ of the
quadratic Lagrangian for fermions:
\eqn\sixdquadr{\eqalign{\int_{W_2} \int_{\cp} dV
& \Biggl({1 \over 2} g^{AC}g^{BD}\p_0
\b_{AB}\bar{\theta}
\Gamma_{0} \Gamma_{C} {\cal D}_D \theta
+{1 \over 4} \epsilon^{ABCD}\p_0
\b_{AB}\bar{\theta}
\Gamma_{1} \Gamma_{C} {\cal D}_D \theta \cr
&- \bar{\theta}
\Gamma^{\alpha}{\cal D}_{\a} \theta
+ {1 \over 4!} \epsilon^{\a \b}
\epsilon^{ABCD}\bar{\theta}
\Gamma_{ABCD}
\Gamma_{\a } {\cal D}_{\b}\theta \Biggr), \cr} }
where $\theta$ is the eleven-dimensional superspace coordinate (the
superpartner of the embedding coordinates $X^M(\sigma)$),
$\Gamma$'s are eleven-dimensional gamma-matrices and ${\cal D}$ is
the pullback of the spin connection from the ambient space to the
fivebrane worldvolume $W_6$. This piece of the action is obtained
by gauge fixing the $\kappa$-symmetric action of
\fiveac\ in a
general curved background, keeping only the terms
quadratic in $\theta$ which
involve exactly one derivative along
$W_2$, and discarding ${\cal O}((X^a)^2)$ terms.
 The last term comes from the coupling of the
six-form potential to the fivebrane worldvolme.

In reducing \sixdquadr\ we can make use of the $\kappa$-symmetry to
eliminate the unphysical degrees of freedom of $\theta$ and express
it in terms of six-dimensional spinors $\psi_{-I}\otimes
\Delta^{I}, \, \chi_-^{I}\otimes (\Delta^*)_{I}$. Here $\psi_{-I},
\, \chi_-^{I}$ are two-dimensional Weyl spinors and $\Delta^I$ are
four-dimensional Weyl spinors (see section 3.2 and appendix B). The
six-dimensional $\Gamma$-matrices decompose as in section 3.2.
 Suppressing
internal (along $\cp$) covariant derivatives on the spinors, the
reduced action
can be cast in the form (the universal superfield is not included)
\eqn\quadrtrms{\int_{W_2} d^2 \sigma \p_+ \varphi^K
[(\psi_{-J})^{\dagger}
\psi_{-I} {\cal R}_K^{ \bar{J}I}
+ (\chi_-^J)^{\dagger} \chi_-^I {\cal R}_{K \bar{J} I}] +{\rm c.c.}
}
where we use the same conventions for the two-dimensional fermions
as in section 3 and we  have defined
\eqn\scpls{
{\cal R}_K^{
\bar{J}I}= \int_{\cp} dV
(\Delta^J)^{\dagger} \nabla_K
\Delta^I
; \,\,\,
{\cal R}_{K
\bar{J}I}= \int_{\cp} dV
(\Delta_{\bar{J}})^{tr} \nabla_K (\Delta^*)_I , }
where $\nabla_I$ is the covariant derivative discussed in section
4.2.

We will now  analyze the meaning of the ``coupling'' of \scpls\ to
extract the target space geometry. As  in   section 4.1  we
consider a family of surfaces $\CP_\varphi$ near $\CP_{\varphi=0}$.
With respect to our basis of holomorphic two-forms $\nabla$ acts in
the following way:
\eqn\omgvar{\nabla_J \omega_I = [\Gamma_J]^x{}_I
\omega_x }
where $x \in \{I, \bar{I}, a \}$ and $[\Gamma_J]$ is the $\varphi$-
dependent
Gauss-Manin connection matrix. Therefore
\eqn\dvartn{\p_K D_{I \bar{J}}=\int_{\cp}\nabla_K \omega_I \wedge
\omega_{\bar{J}}= \int_{\cp} [\Gamma_K]^L{}_I \omega_L \wedge
\omega_{\bar{J}},}
where we have defined $\p_I ={\p \over \p \varphi^I}$.
On  the other hand
$\Delta_I=\omega_I(\gamma^{(2)})\xi$ , $\,
\gamma^{(2)}=
\gamma^A
\gamma^B {\p \over \p x^A} \otimes {\p \over \p x^B}$
and it's easy to see that
\eqn\rhoeqd{\eqalign{{\cal R}_{K \bar{J}I}=
\int_{\cp}dV \xi^{\dagger}
\omega_{\bar{J}}&(\gamma^{(2)}) \nabla_K \omega_I(\gamma^{(2)})\xi \cr
&=\int_{\cp} dV \xi^{\dagger}
\gamma ^{\bar{m} \bar{n}} \omega_{\bar{J}\bar{m} \bar{n}}
[\Gamma_K]^x{}_I \omega_{xAB} \gamma^{AB} \xi =\p_K D_{I \bar{J}}.
\cr}}
For the last step we have used \dvartn, the fact that
$\gamma_{\bar{m}}
\xi =0$ (as in 4.3) and a little bit of gamma-matrix algebra.
We conclude that ${\cal R}_{K
\bar{J}I}$ is just the Christoffel symbol of the manifold
$|\cp|$ with  K\"{a}hler metric \mercf.

\subsec{Comparison to the standard $(0,4)$ Lagrangian}

Let us now assemble the data we have gathered and compare
to the standard Lagrangian  spelled out in section two.
The part of (the bosonized version of) the $(0,4)$ action
\compaction\ containing all one-derivative terms quadratic in
fermions is
\eqn\bosoni{ \psi^i_- \psi^j_- [F_{ij{\hat a}} \pap \rho^{\hat a}
+ \Upsilon^{(+)}_{ijk} \pap \phi^k].}
where $\rho^{\hat a}$, $\hat a = 1,\dots, b_2^- -b_2^+$
 is the set of
purely left-moving scalar fields.

Comparison to \quadrtrms\ implies that the $b$-field and the gauge
connection of
the vector bundle over the target-space are flat, and that the metric on
the
target-space is
\eqn\trgtspmtrc{ds^2=D(\varphi,\bar{\varphi})_{I\bar{J}}d\varphi^I
d\bar{\varphi}^{\bar J}
        +D(\varphi,\bar{\varphi})^{I\bar{J}}(\nabla {\bar \pi})_I (\nabla
\pi)_{\bar{J}}  }
where $D_{I\bar{J}}$ is the intersection pairing defined in \mercf.
Since the connection on $T\cmt$ has no torsion supersymmetry
requires that the metric \trgtspmtrc\ is hyper-K\"ahler, and
indeed, gives a way
 to prove the hyperk\"ahler property of the metric.
 Using the supersymmetry transformations \rdftf\
and the fact that $(0,4)$ symmetry is unbroken, it follows that
the 3 complex structures in \cplxstrs\
are covariantly constant.
\foot{It would be desirable to have a more direct,
and more standard proof of this fact. This is being
investigated in \mmtt.}

It is interesting to compare the metric with the
 c-map construction \cfg. The metric there
reads
\eqn\cmpmtr{ds^2=D_{I\bar{J}}d\varphi^I \otimes
d{\bar \varphi}^{\bar{J}}+D^{I\bar{J}} (\nabla {\bar \pi})_I
\otimes (\nabla
\pi)_{\bar{J}} }
(note that in this case $D_{I\bar{J}}=D_{\bar{I}J}$) One has three closed
two-forms $\omega^1,
\,
\omega^2,
\,
\omega^3$ where $\omega^1$ is the associated $(1,1)$ form and
$\omega^{1,2}$ are the real, imaginary parts of $d\varphi^I \wedge
(\nabla {\bar \pi})_I$. Setting $d\varphi^I= d\phi^{1I}+
id\phi^{2I}; (\nabla \pi)^I:= D^{I \bar{J}} (\nabla
\pi)_{\bar{J}}= d\phi^{3I}+ id\phi^{4I}$, the metric takes the form
$ds^2=G_{iI,jJ}d\phi^{iI}
\otimes d\phi^{jJ}$, where $G_{iI,jJ}=D_{I \bar{J}} \delta_{ij}$. The
three two-forms can be used to construct three complex structures
$J_r{}^{iI}{}_{jJ}:=G^{iI,kK}\omega^r_{kK,jJ}$ which in components read:
$J_1=[\sigma^3 \otimes i\sigma^2]^i{}_j \delta^I{}_J, \,\, J_2=[i\sigma^2
\otimes
\I1_2]^i{}_j \delta^I{}_J,
\,\, J_3=[\sigma^1 \otimes i\sigma^2]^i{}_j \delta^I{}_J$. These are
exactly the same
as in \cplxstrs.

\newsec{Comments on the global structure of the target space}

\subsec{Narain theory in the entropic factor}

We now consider to the periodicities of $\beta$, needed to determine the
global structure of the fibers of $p$ in
\intsys. As we have stressed in the introduction,
we expect the target space to be compact. Therefore, while the
local target manifold is
 $\IR^4 \times T^*\vert P\vert $, the fibers should be compactified,
maintaining the hyper-K\"ahler property. The most natural (perhaps
the only) way to do this is to take a quotient by a lattice in the
fiber of $T^*\vert \CP\vert $ so that
 the fibers of $\cmt$  are complex tori.

Passing to a real basis $\{
\omega_{\hat{I}}, {\hat{I}}=1,
\dots b_2^+ \}$ of self-dual two forms on $\cp$ we can expand
\eqn\bdcmpa{\beta = \pi^{\hat{I}} \omega_{\hat{I}} + \rho^a \omega_a}
and reexpress \fixedact\ as
\eqn\fixedacta{S_0 = \int_{W_2}d^2 \sigma ( \pao \pi^{\hat{I}}  \pap
\pi^{\hat{J}}
 D_{{\hat{I}}{\hat{J}}} - \pao \rho^a \pam \rho^b D_{ab}).}
where $D_{{\hat{I}}{\hat{J}}}= \int_{\cp}\omega_{\hat{I}}
\wedge
\omega_{\hat{J}}$.
Thus the metric is diagonal on left- and right-movers.
However, the information of how
to ``combine'' the zero-modes of the left and a right-moving bosons to
define the
statespace of the full conformal field theory
is not contained
in \fixedacta.
This has to be imposed {\it  ad hoc}.

Our ansatz is that the Lagrangian in \fixedacta\ is
 a Narain $\sigma$ - model
with non-trivial Narain data: a constant metric, a constant torsion,
and Wilson lines. Thus, we take the periodicities
\eqn\beeperiod{
\beta
\rightarrow \beta + n^{\rm x} U_{\rm x}, \,\, n^{\rm x} \in
\IZ
}
where we have introduced a basis
$\{ U_{\rm x}; \,\,\, {\rm x}=1, \dots b_2 \}$ of $H^2(
\cp , \IZ)$.
The data for the zeromodes of the scalars
 are encoded in
the projections onto the definite signature subspaces:
\eqn\narain{P: H^2(\cp;\IZ) \otimes \IR
\rightarrow  H^{2-}(\cp; \IR) \perp H^{2+}(\cp; \IR).
}
In particular, the left and right-moving momenta are
just  ${\bf p}=(F^a_{\rm x} n^{\rm x} {\bf
e}_a;\,f^{\hat{I}}_{\rm x} n^{\rm x}{\bf e}_{\hat{I}})$, where
$F^a_{\rm x} =
\int_{\cp}\omega_a
\wedge U_{\rm x}; \,$ $f^{\hat{I}}_{\rm x}
 = \int_{\cp}\omega_{\hat{I}} \wedge U_{\rm x}$,
and ${\bf e}_a$ is the vielbein for the metric $D_{ab}$ and ${\bf
e}_{\hat{I}}$ is the vielbein for $D_{{\hat{I}}{\hat{J}}}$.

\subsec{Charge violation by instantons: The ``MSW effect''}

The Narain model of the previous section is
somewhat peculiar because the conserved
$U(1)$ charges coupling to the string are
in  the
lattice $H^2(X;\IZ)$ which is a
(small!) sublattice of $H^2(\CP;\IZ)$.
This puzzle was resolved in
\msw\ as follows.
\foot{We thank J. Maldacena and E. Witten
for important clarifying explanations
about this process.}
The charges $H^2(\CP;\IZ)$ are conserved in
the $(0,4)$ sigma model studied in this
paper, but they are violated by
membrane instanton processes in the
full $M$-theory.
As  mentioned in \msw\ if a
state in the $(0,4)$ CFT is charged under
an element in $H^2(\CP;\IZ)$ which is
not in $H^2(X;\IZ)$ it will decay to
a state charged in $H^2(X;\IZ)$.
Indeed, since the map \restr\ is
injective the dual map:
\eqn\surjctv{
H_2(\CP;\IZ)
~ {\buildrel \iota_* \over \rightarrow}~ H_2(X;\IZ) \rightarrow 0
}
is surjective, and hence has a large kernel.
Elements of the kernel are nontrivial
surfaces $[\Sigma] \in H_2(\CP;\IZ)$ which
bound a 3-ball in $X$, $\Sigma = \p B$.
It is possible to have a
 membrane instanton whose worldvolume is
$B$ because  the equation
\eqn\sourceq{
d H = -Q(M2) \delta(\Sigma \hookrightarrow W_6), } where $H=d\b$
and $Q(M2)$ is the membrane charge, allows membranes to end on
fivebranes
\open. Since this process uses $M$-theory instantons, it will only
be important near degenerations of $\CP$.

One interesting question raised by  this  ``MSW effect''
is whether states on the 5brane can carry
torsion charges. The kernel of $\iota_*$ is a
sublattice of $H_2(\CP;\IZ)$.
We claim that under
 Poincar\'e duality $PD: H_2(\CP;\IZ)
\rightarrow H^2(\CP;\IZ)$, we have
$PD(\ker \iota_*) = \iota^*(H^2(X;\IZ))^\perp$
where the orthogonal complement is in the
Hodge metric of $\CP$.
 To prove this note that if
$[\Sigma]\in \ker\iota_*$ then its Poincar\'e
dual form $\eta_\Sigma\in H^2(\CP;\IZ)$
satisfies
\eqn\orthog{
\int_{\CP} \eta_\Sigma \wedge \iota^*(\theta)
= \int_{\iota_*(\Sigma)} \theta
} for all $\theta\in H^2(X;\IZ)$. Since $\iota^*(H^2(X;\IZ))$ is
not unimodular while $H^2(\CP;\IZ)$ is unimodular, the sublattice
$\iota^*(H^2(X;\IZ)) \oplus \iota^*(H^2(X;\IZ))^\perp$ will have
finite index in $H^2(\CP;\IZ)$. The quotient group is a (large)
group of potential torsion charges. We say ``potential'' because we
do not fully understand the model globally on $\vert P\vert$. It
would be interesting to understand how the above torsion charges
can be understood in the framework of the K-theory interpretation
of D-brane charges \refs{\mm,  \wittenk}.

\subsec{Narain data for the universal factor }

It is possible to be much more explicit about the Lagrangian for
the universal multiplet. Just as  for the rest of the fields, its
action follows from the reduction of \simplesch\ and \sixdquadr.

Let $\{ J, \, \theta_{\Lambda}; \, \Lambda =1, \dots h^{1,1}(X)-1
\}$ be a basis of $H^{1,1}(X,\IR)$ such that $\theta_{\Lambda}$
restricts to a basis of anti-self-dual forms on $\CP$. Moreover,
let
 $\{ Y_w; \, w=0, \dots h^{1,1}(X)-1 \}$ be a basis of $H^{1,1}(X,\IZ)$. The
part  $\b_{u}$ of the chiral 2-form
 contributing to the ``universal''
multiplet is expanded as
\eqn\bunexp{\b_{u}=u_4 J+ \rho^{\Lambda} \theta_{\Lambda}}
with the periodicities
\eqn\bunper{\b_{u} \rightarrow \b_{u}+ n^w Y_w; \,\,\,n^w \in \IZ.}
The universal multiplet is governed by the action
\eqn\univact{S_{un}=\int_{W_2}d^2 \sigma \{ D_{00} \p_0 u_4 \p_+ u_4
- D_{\Lambda \Lambda'} \p_0 \rho^{\Lambda} \p_- \rho^{\Lambda'} \} }
where $D_{\Lambda \Lambda'}:=\int_{\cp} \theta_{\Lambda} \wedge
\theta_{\Lambda'}
= \int_X P
\wedge \theta_{\Lambda} \wedge
\theta_{\Lambda'} ;
\,\, D_{00}:=\int_{\cp} J \wedge J=2Vol(\cp)$.

Repeating the analysis of section 6.1 we see that the left/right-moving
momenta
are given by
\eqn\lrmm{{\bf p}=(F^{\Lambda}_w n^w {\bf e}_{\Lambda}; \, f^0_w n^w {\bf
e}_0)}
where $F^{\Lambda}_w :=\int_{\cp} \theta_{\Lambda}\wedge Y_w; \,
f^0_w
:=\int_{\cp} J
\wedge
Y_w$ are the projections
\eqn\prj{P: \, H^2(X, \IZ) \otimes \IR \rightarrow H^{2+}(X, \IR) \oplus
H^{2-}(X, \IR),}
and ${\bf e}_{\Lambda} \, ({\bf e}_0)$ is a vielbein for the
metric $D_{{\Lambda}{\Lambda}'} \, (D_{00})$. The self-dual
(right-moving) piece is generated by $J$. The the radius $R$ of the
$S^1$ in the target space is given by
 $R^2={1\over 2\pi^2}Vol(\cp)$, when
$h^{1,1}(X)=1$, and by more complicated formulae
in general.

\subsec{Effects of the $M$-theory 3-form}

Finally, let us comment on two effects
that happen  when we
  turn on the $C_3$ field of the
eleven-dimensional supergravity.

First, in the Kaluza-Klein reduction of
$M$ theory on $X$ the field $C_3$
 gives rise to $h^{1,1}(X)$
five-dimensional vectors (together with KK modes from the metric
these form the gravity multiplet and  $h^{1,1}(X)-1$ vector
multiplets). The coupling of $C_3$ to the fivebrane worldvolume
induces string couplings to the background gauge fields
\eqn\gauco{\int_{W_2} d^2\sigma \{ A^{\Lambda}_+
\pam \rho^{\Lambda'} D_{\Lambda \Lambda'} +
A^0_- \pap u_4 D_{00} \} }
where $A^{\Lambda}$ are the abelian vector fields and $A^0$ is the
graviphoton. Such couplings are also important for cancellation of
anomalies in the gauge transformations in the presence of the
string \fkm. Since the projections in
\prj\ already encode Narain data, including the flat connection on
the gauge bundle, we see that turning on $C_3$ just shifts the
gauge fields.

Second, the 5brane action consists of a Dirac part and a WZ part.
In the Dirac part the fieldstrength of the chiral 2-form enters
through $H= d \beta - C_3$. In the Kaluza-Klein reduction this
leads to shifts of the periodicities of the chiral scalars, for
example, $\p_+ \pi^{\hat I} \rightarrow \p_+ \pi^{\hat I} + C^{\hat
I}$.  If $C_3 = dX^1 \wedge \theta$, with $\theta \in H^2(X;\IR)$
then the  Narain vectors are shifted by $p \rightarrow p + \theta$,
leading to a shift in $L_0 - \tilde L_0$. If we consider the
corresponding IIA picture this  is in accord with
the  Witten-effect shifting of the
 D0 charge:
\eqn\shift{
\Delta(L_0 - \tilde L_0) =
\int_{\CP} \biggl( p\wedge \theta + \half \theta \wedge \theta \biggr).
}
where we have identified $p$ with the first Chern class of the
Chan-Paton bundle on the D4 brane.

\newsec{Conclusion: 5  problems on 5 branes}

First and foremost it would be good to extend the
discussion in this paper to understand the global
geometry on $\vert P \vert$. This consists of
at least two important sub-problems.
First, we have restricted to an open neighborhood
in $\vps$. It would be interesting to take into
account the effects of monodromy.
Second,
the 4-cycle $\CP$ will degenerate on a codimension one
discriminant locus $\CD= \vert
P\vert - \vert P\vert_{\rm s}$ of the linear system.
The generic singularity will be a rational double point.  Many
interesting and important questions depend crucially on
understanding what happens to the $(0,4)$ model when the fivebrane
degenerates. In \maldacena\ a drastic degeneration with $D$ points
of self-intersection was successfully used to count black hole
entropy at leading order in large charges.

Second, as mentioned in the
introduction, one of the original
motivations for this work was to find a
state-counting formula for BPS states in $M$-theory
compactifications which are macroscopically
4d black holes with $8$ supersymmetries.
We believe that combining the elliptic genus of
$(0,4)$ models studied in \kawaimohri\
with the results of this paper one can derive
formulae for the BPS degeneracies. This idea
is currently under investigation.

Third, it would be nice to clarify the status of the above model as
a CFT. Since the $\sigma$-model
described in this paper is rather elaborate, it would
be nice to have a clear understanding of
whether  the entropic factor
is, in fact, a conformal field theory (and if not, what it flows to).
Moreover,  it might be useful to find a linear sigma
model which renormalizes to the above nonlinear model. This would
be possible if the metric on $T^*\vert P \vert$ were given by a
hyper-K\"ahler quotient. Thus, an
interesting question raised by this work is whether
there is a sense in which the metric on $T^*\vert P\vert$ induced
by the Calabi-Yau metric becomes the hyperk\"ahler quotient
metric in the limit of large $P$.

Fourth, it would be nice to extend the discussion to fivebranes
with even less supersymmetry, leaving a $(0,2)$ string. Such
configurations would appear if the $M5$ worldvolume is near a
boundary, as in the Horava-Witten picture. At a formal level, much
of the above discussion generalizes to the $(0,2)$ case. However
quantum corrections are expected to be much more important here.

Fifth , if the $M$-theory compactification has a heterotic dual
then there must be a description of the same strings in the
heterotic picture. Indeed, in the case $X = K3 \times T^2$ with
$\CP = K3$ one reproduces the heterotic string itself \refs{\hast,
\cs}. However, in the case of $\CP$ defined by a class $P$ with $P$
large there will be a large number of left- and right-moving
degrees of freedom. Because of the MSW effect it is not obvious
that these charges should really be visible. We think this is worth
understanding better.

\bigskip
\centerline{\bf Acknowledgments}\nobreak
\bigskip

GM would like to thank J. Maldacena and E. Witten for several
important discussions on this subject. We would also  like to thank
P. Deligne, D. Freed,  D. Morrison, T. Pantev, A. Todorov, and G.
Zuckerman for discussions.  RM acknowledges the hospitality of the
Erwin Schr\"{o}dinger Instutute, Ecole Polytechnique and LTPHE,
Paris VI-VII. GM thanks the Institute for Advanced Study for
hospitality and the Monell Foundation for support during the
completion of this paper. This work is also supported by DOE grant
DE-FG02-92ER40704.

\appendix{A}{List of some notation}

\subsec{General notation}

$A_{[1\cdots k]} = A_{1\cdots k}$ for a $k$-form $A$

$\alpha= 0,1$: the directions along the string world-sheet

$a = 6,...10 $ : the directions transverse to $W_6$;

($a=1,...b_2^-$ also enumerates the basis vectors of $H^{2-}(\cp )$)

$\hat{a}= 1, \dots b_2^- -b_2^+$

$A, B = 2,...5 $ : the (real) directions along ${\cal P}$.

$\beta$ \qquad The chiral 2-form of the 5brane 6D tensormultiplet.

$\gamma^{\mu}, \, \Gamma^{M}, \,  \hat{\gamma}^a$: Gamma matrices
defined in sections 3, 4.3 and B1.

$\check{\gamma}^1 - \check{\gamma}^4$: matrices defined in section
4.4 (below \transun).

$D_{I \bar{J}} \equiv \int_{\cp} \omega_I \wedge \omega_{\bar{J}}$

$D_{ab} \equiv \int_{\cp} \omega_a \wedge \omega_{b}$

$H^{1,1}(X)^{\perp}$: The subspace of $H^{1,1}(X)$ ``orthogonal''
to $J$, see section 4.1.

$H^{2 \pm}(\cp)$: the spaces of self-dual, antiself-dual 2-forms
on $\cp$.

$\theta_{\Lambda}, \, \Lambda=1,...h^{1,1}(X)-1$: a basis of
$H^{1,1}(X)^{\perp}$.

$i=1,...4$: a $USp(4)$ index, except in section 2.

$I=1,...{1 \over 2} (b^2_+ -1)$, except in section 2.

${{{\hat{I}}}}=1, \dots b_2^+$

$J$: the K\"{a}hler form on $\cp$ and on $X$

$\cl$: the holomorphic line bundle over $X$, associated to the
divisor $\cp$.

$M=0,...10 $: the spacetime index

$\mu, \nu = 0,...5$ : the directions along $W_6$.

$m, \bar{m} = 1,2$ : the (complex) directions along ${\cal P}$.

$N$ \qquad The number of $(0,4)$ multiplets. Defined in \dimels.

$\omega_{-a}, \,\, a=1,...b_2^-$: a basis of $H^{2-}(\cp )$.

$\omega_I \, (\omega_{\bar{I}}) , \, I=1,...{1 \over 2} (b^2_+
-1)$: A basis of $H^{(2,0)}(\cp)$ ($H^{(0,2)}(\cp )$)

$\omega_{{{\hat{I}}}}, \, {{{\hat{I}}}}=1, \dots b_2^+$: a basis of
$H^{2+}(\cp ,\IR)$

$\CP$: \qquad A generic smooth holomorphic surface inside $X$.

$\vert P\vert_{\rm s}$ \qquad The locus of smooth divisors in
the linear system $\vert P\vert$.

$P$: \qquad The cohomology class in $H^2(X;\IZ)$ dual to the 4-cycle
$\CP$.

$\sigma^0 - \sigma^5$: the coordinates on $W_6$

$U_{\rm x}, \,\, {\rm x=1}, \dots b_2(\cp)$: a basis of $H^2(\cp ,
\IZ)$

$Y_{\rm w}, \,\, {\rm w=1}, \dots h^{1,1}(X)$: a basis of
$H^{1,1}(X ,\IZ)$

$X$ \qquad A Calabi-Yau 3-fold, used for compactifying $M$-theory.

$X^M( \sigma )$ : the embedding of $W_6$ to the eleven-dimensional
spacetime.

$\xi^{(6)}$: the covariantly constant spinor of the Calabi-Yau.

$\xi$: the component of $\xi^{(6)}$ along $\cp$ (in a local
decomposition).

\subsec{Conventions for Section 5.1 }

$\hat{ \mu }, \hat{ \nu } = 0,...5 $ : the directions along $W_6$.

$ \mu, \nu = 0,2,...5$ : omitting the ``distinguished'' direction.

$\sigma^{ \hat { \mu } }$ : the coordinates on $W_6$.

$G_{ \hat{\mu} \hat{ \nu} } = \eta_{ MN } \p_{ \hat{\mu} } X^M \p_{
\hat{ \nu} } X^N$

$G_5 =  det(G_{ \mu \nu }) $

$H_{\mu \nu \rho} = 3 \p_{ [ \mu } \beta_{ \nu \rho ] }$

$\tilde{ H }^{\mu \nu } = {1 \over 6} \varepsilon^{ \mu \nu \rho
\kappa \lambda } H_{ \rho \kappa \lambda } $

\appendix{B}{$(2,0)$ tensor multiplet }
\subsec{The conventions}
In this section we work in six-dimensional Minkowski space. The
$R$-symmetry group for the theory with sixteen real supercharges is
$SO(5)$.  Let $a=1,..., 5$ (the index $a$ is $SO(5)$ Euclidean) and
$\mu=0,1,...5$.
A basis of gamma-matrices $\hat{\gamma}^{a}$ in five-dimensional
Euclidean space  can be constructed as follows:
\eqn\fivegammabasis{ \hat{\gamma}^{6,7,8} = \pmatrix{
0 & \sigma^{1,2,3} \cr
\sigma^{1,2,3} & 0 \cr
}; \,\,\,\,\,  \hat{\gamma}^{9} = \pmatrix{
0 & i \, \I1_2 \cr
-i \, \I1_2 & 0 \cr
}; \,\,\,\,\,
\hat{\gamma}^{10} = \pmatrix{
- \I1_2 & 0 \cr
0 & \I1_2 \cr
} }
In checking the susy transformations of the (2,0) multiplet of
$B.2$ it is more convenient to work in a slightly different basis
than the one we used in $4.3$ for gamma-matrices $\Gamma^{\mu}$  in
six-dimensional Minkowski space:
\eqn\sixgammabasis{\Gamma^{\mu}=\pmatrix{
0 & \gamma^{\mu} \cr
\tilde{\gamma}^{\mu} & 0 \cr
}}
where
 \eqn\gms{\eqalign{\gamma^0=\pmatrix{
0 & \I1_2 \cr
-\I1_2 & 0 \cr
}; \,&\,\,\,\, \gamma^{1,2,3}= \pmatrix{
0 & \sigma^{1,2,3} \cr
\sigma^{1,2,3} & 0 \cr
}; \,\,\,\,\,\, \gamma^4=\pmatrix{
\I1_2 & 0 \cr
0 & -\I1_2 \cr
} \cr
&\tilde{\gamma}^{0-4} = \gamma^{0-4}; \,\,\,\, \tilde{\gamma}^5=
-\gamma^5= -i \, \I1_4  \cr
}}

In this basis the charge-conjugation and the chirality matrices are
\eqn\ccach{{\cal C}^{(6)}=\pmatrix{
0 & c \cr
-c & 0 \cr
};\,\,\,\,\,\,\,\rho^{(6)}=i^2\Gamma^1... \Gamma^5 \Gamma^0=\pmatrix{
\I1_4 & 0 \cr
0 & -\I1_4 \cr
}}
where
\eqn\defc{c=\pmatrix{
\epsilon & 0 \cr
0 & -\epsilon \cr
}}
with $\epsilon$ given by
\eqn\epsilons{\epsilon_{AB}=\pmatrix{
0 & 1 \cr
-1 & 0 \cr
}=i\sigma^2; \,\,\, \epsilon^{AB}=-\epsilon_{AB}.}
The real, antisymmetric tensor of $USp(4)$ obeys
\eqn\wmega{\Omega=-\Omega^{-1}; \,\,\,\,\,\,\,
\Omega (\hat{\gamma^{a}})^{Tr}=
\hat{\gamma}^{a}\Omega}
We can write $\Omega$ in this basis explicitly
\eqn\explqm{\Omega=\pmatrix{
-\epsilon & 0 \cr
0 & \epsilon \cr }} For an (anti)chiral Spin(1,5) spinor
$\theta_i,\,\, i=1,...4$ transforming in the $\bf{4}$ of $USp(4)$
($i$ is a $USp(4)$ index) the symplectic-reality condition reads
\eqn\sympreal{\theta_i=\Omega_{ij}c\bar{\theta}^{Tr}_j;
\,\,\,\,\,\,\,\,\bar{\theta}_i
=-\Omega_{ij}\theta_j^{Tr}c}
where $\bar{\theta}=\theta^{+}\gamma^0$.

\subsec{$6D$  supersymmetry}

 The 6D $(2,0)$ multiplet consists of a self-dual
(on-shell) antisymmetric two-form $\b_{\mu\nu}$, four
six-dimensional Weyl spinors $\{
\psi_i,
\, i=1
\dots 4 \}$ obeying the symplectic reality condition (B.9),
and five scalars $\{ X^a, \, a=6,...10 \}$. In other words, under
the little group $Spin(4) \times USp(4) \,$, $\b_{\mu \nu}, \,
\psi, \, X^a$ transform in the
$((\bf{3},\bf{1});\bf{1}), \, (\bf{4};\bf{4}), \,
((\bf{1},\bf{1});\bf{5})$ respectively. After the elimination of
the auxiliary field introduced in the covariant formulation of the
fivebrane of \fiveac, supersymmetry closes on-shell.

The susy transformations are (suppressing the USp(4) index on the
fermions):
\eqn\sixsusy{\eqalign{
\delta X^{a} &= -2{\bar \epsilon} \hat{\gamma}^a \psi \cr
\delta \psi &=\left(  \half \gamma^{\mu} \partial_{\mu}X^a \hat{\gamma}_a
+
{1 \over 8}
\gamma^{\mu\nu\rho} H_{\mu\nu\rho} \right)\epsilon \cr
\delta \beta_{\mu\nu} &= -2{\bar \epsilon} \gamma_{\mu\nu} \psi \cr}}
where $H_{\mu\nu\rho}= \partial_{[\mu} \beta_{\nu\varrho]}$. Using
these equations one checks that the algebra closes on shell.

\appendix{C}{Some remarks on Kodaira-Spencer theory}

In this appendix we show \eqvdfs. Let $X$ be
a complex manifold with a divisor ${\cal P}$. There are two exact
sheaf sequences which we are going to use
\eqn\shfsca{0 \rightarrow {\co(T \cp)} \rightarrow {\co(TX|_{\cp})}
\rightarrow
{\co(\cl|_{\cp})} \rightarrow 0,}
which is a sequence over $\cp$, and
\eqn\shfscb{0 \rightarrow \co(TX \otimes [- \cp]) \rightarrow \co(TX)
\rightarrow  \co(TX|_{\cp}) \rightarrow 0,}
which is a sequence over $X$. From
the long exact sheaf-cohomology sequence associated to \shfsca\
we obtain:
\eqn\lnga{\cdots \rightarrow  H^0( \cp,\co(TX|_{\cp})  \rightarrow
H^0(\cp,
\co(\cl)) \rightarrow H^1(\cp, \co(T \cp))
 \rightarrow H^1( \cp,\co(TX|_{\cp})) \rightarrow
\cdots}
Since $ H^1(\cp, \co(T \cp)) \cong H^{0,1}(T\cp),$ in order to show that
the
mapping \eqvdfs\ is injective, it suffices to show  that $H^0(
\cp,\co(TX|_{\cp})) = 0$. For this we will use the following part of the
exact
long sheaf-cohomology sequence associated to \shfscb:
\eqn\lngb{\cdots \rightarrow  H^0( X,\co(TX))  \rightarrow H^0(\cp,
\co(TX|_{\cp})) \rightarrow H^1(X, \co(TX \otimes [-\cp])) \rightarrow
\cdots,}
where we noted that $H^*(X,\co(TX|_{\cp})) = H^*(\cp,
\co(TX|_{\cp}))$. However  $H^0( X,\co(TX) )\cong H^{0,0} (TX)
\cong H^{2,0}(X) =0$ (where the last equivalence can be seen using
 the existence of
a unique nowhere-vanishing holomorphic three-form on $X$). Moreover
using Kodaira-Serre duality we have
 \eqn\kdsr{H^q(X, \co(TX \otimes [-\cp])) \cong H^{3-q}(X,\Omega^1(\cl))^*
=0;
 \,\,\,\, q=0,1,2,}
where the last equality is due to the fact that $\cl$ is associated
to the very ample divisor $\cp$, and we can take $c_1(\cl)$ to be
arbitrarily large. Thus it immediately follows from \lngb\ that
$H^0(\cp, \co(TX|_{\cp})) = 0$ and hence \eqvdfs\ is indeed
injective. One can show that
$H^1(\cp,\co(TX|_{\cp}))
\cong H^1( X,\co(TX)) \cong H^{0,1} (TX) \cong H^{2,1}(X) \neq 0 $
so we cannot
conclude from \lnga\ that \eqvdfs\ is surjective.

\appendix{D}{D3 on K3 and the $(4,4)$ $\sigma$-model}
Although outside the main line of development
of this paper, it is worthwhile discussing
 the properties of $(4,4)$ models within the
  framework of this paper. For a recent account see
\rd. Here we address some complementary issues. To obtain a $(4,4)$
model we will consider a $D3$ wrapped on a holomorphic two-cycle (a
Riemann surface) $\cal P$ inside $X=K3$ (whenever it doesn't lead
to  confusion, we will keep the same notation as for the
corresponding discussion in the case where X is a Calabi-Yau
three-fold). This is a much simpler system to analyze, since all
the scalars coming from the reduction to the string world-sheet of
the $D3$-brane low-energy lagrangian, are non-chiral. The number of
left and right-movers is given by a formula similar to the one for
the fivebrane:
\eqn\dcount{N_L^B = N_R^B = d_P + b_1(\cp) + 4}
The (bosonic part of the) gauge theory on the worldvolume involves
a vector field $A_{\mu}$ and six scalars. When wrapped on the
two-cycle, two scalars $X^4, X^5$ will parametrise the
deformations, yielding $d_P$ scalars on the string worldvolume,
while the other four $X^6-X^9$ will form the universal superfield
(here, in analogy to the discussion for the $M5$, we consider the
$D3$-brane to be along $X^0-X^3$ while X is taken along $X^2-X^5$).
In this case, the universal superfield does not contain compact
scalars, and is given simply by $\IR^4$. In its turn, the vector
field gives rise to $b_1(\cp)$ scalars. Note that the K\"{a}hler
form no longer appears in our analysis of the scalar spectrum and
the structure of the universal superfield is considerably simpler.
More precisely, the counting goes as follows
\eqn\hip{\chi(\CP) = \int_{cp}c_1(\cp) = - \int_{X} P^2 = 2 - b_1(\cp),}
where as before we have set $P = c_1(\cl) = -c_1(\cp)$. On the
other hand
\eqn\hil{\chi(\cl)= \sum_{i=0}^{dimX}(-1)^i {h^i(X, \cl)}= h^0(X, \cl),
\,\,\,\, h^i(X, \cl)\equiv dim_{\IC} H^i(X, \cl) }
where the last equality follows from the fact that $\cp$ is very
ample, and
\eqn\indxthm{ h^0(X, \cl)= \int_X{e^P Td(X)} =
\int_X{ \left( {1 \over 2} P^2 + {1 \over 12} c_2(X) \right)}. }
Taking \hip\ into account and the fact that $\chi(X)=
\int_X {c_2(X)} =24$ for $X=K3$, we finally get
\eqn\rldm{d_P=b_1(\cp)=2D + 2,}
where again $d_P$ stands for the real dimension of $H^0(X, \cl)$
and $D={1 \over 2} \int_X P^2$ as before. The $(4,4)$ action
follows from the reduction of the Born-Infeld action for $D3$
\eqn\llead{L= -{1 \over 4}F^2 +
\sum_{a=4}^9{\p_{\alpha}X^a \p^{\alpha}X^a}+ \dots  }
Let $\{\omega_I(\varphi), \,\,\, I=1, \ldots, {1 \over 2} b_1 \}$
 be a basis of
holomorphic one-forms on $\cp_{\varphi}$. We can write the gauge
field in terms of this basis
\eqn\gfexp{A^m= \pi_I \omega^{I, m} }
Moreover, to the first order in $\varphi$ we expand
\eqn\scexpn{X^4+iX^5 = \varphi^I \upsilon_I,}
with $\{ \upsilon_I \}$ a basis of holomorphic sections of
$\cl_{\cp}$.

The $c$-map  works in the same way as in the  case of \cfg, and we
obtain the result that all the terms in the reduction of
\llead\ come from a K\"{a}hler potential of the form \foot{To prove
this, one should use $\omega_I
= \iota(\upsilon_I) \Omega \vert_{\CP}^{(2,0)}$.}
\eqn\khlrptd{ \tilde{ \kappa} = \kappa(\varphi, \bar{\varphi})+
 D^{I \bar{J}} \pi_I \bar{\pi}_{J}; \,\,D_{I \bar{J}}=
 \p_I \p_{\bar{J}}\kappa,}
Again supersymmetry mixes the scalars effectively doubling the
coordinates and yielding a target space of real dimension $2 b_1=
4D+4$.

Finally, the coupling of $D3$ to background RR fields gives rise to
the $\sigma$-model $b$-field. In particular we have
$\int_{D3}\phi^{\rm RR} F \wedge F,$ where the RR scalar has to be
kept fixed in the $D3$ background. The discussion of the
compactness of the target space as well as the dependence of the
$\sigma$-model data in terms of $K3$ geometry follows the generic
Calabi-Yau constructions.

\listrefs
\end